\documentclass[useAMS,usenatbib,aas_macros]{mn2e}
\usepackage{color}

\usepackage{graphicx,graphics}
\usepackage{amsmath}

\title[Stellar mergers in globular clusters]{The possible role of stellar mergers for the formation of multiple stellar populations in globular clusters}
\author[Long Wang et al.]{Long Wang$^{1,2,3}$\thanks{E-mail: longw@uni-bonn.de; }, Pavel Kroupa$^{2,4}$, Koh Takahashi$^{5}$ and Tereza Jerabkova$^{2,4,6}$ \\
$^{1}$ Argelander Institut F\"ur Astronomie, Auf Dem H\"ugel 71, 53121, Bonn, Germany \\
$^{2}$ Helmholtz-Institut f\"ur Strahlen- und Kernphysik, University of Bonn, Nussallee 14-16, D-53115 Bonn, Germany \\
$^{3}$ RIKEN Center for Computational Science, 7-1-26 Minatojima-minami-machi, Chuo-ku, Kobe, Hyogo 650-0047, Japan\\
$^{4}$ Charles University in Prague, Faculty of Mathematics and Physics, Astronomical Institute,\\
V Hole\v{s}ovi\v{c}k\v{a}ch 2, CZ-180 00, Praha 8, Czech Republic \\
$^{5}$ Max-Planck-Institut f\"ur Gravitationsphysik (Albert-Einstein-Institute), Am M\"uhlenberg 1, \\
D-14476, Potsdam-Golm, Germany \\
$^{6}$ European Southern Observatory, Karl-Schwarzschild-Str. 2, 85748 Garching, Gernamy
}

\begin{document}

\date{Accepted --.  Received --; in original form --}

\pagerange{\pageref{firstpage}--\pageref{lastpage}} \pubyear{2002}

\maketitle

\label{firstpage}

\begin{abstract}
  Many possible scenarios for the formation of multiple stellar populations (MSP) in globular clusters (GCs) have been discussed so far,
  including the involvement of asymptotic giant branch stars, fast rotating main sequence stars, very massive main sequence stars and mass-transferring massive binaries based on stellar evolution modelling.
  But self-consistent, dynamical simulations of very young GCs are usually not considered.
  In this work, we perform direct $N$-body modelling such systems with total masses up to $3.2\times10^5$~M$_\odot$, taking into account the observationally constrained primordial binary properties, and discuss the stellar-mergers driven both by binary stellar evolution and dynamical evolution of GCs.
  The occurrence of stellar mergers is enhanced significantly in binary-rich clusters such that stars forming from the gas polluted by mergers-driven ejection/winds would appears as MSPs.
  We thus emphasize that stellar mergers can be an important process that connects MSP formation with star cluster dynamics, and that multiple MSP formation channels can naturally work together.
  The scenario studied here, also in view of a possible top-heavy IMF, may be particularly relevant for explaining the high mass fraction of MSPs (the mass budget problem) and the absence of MSPs in young and low-mass star clusters.
\end{abstract}


\section{Introduction}

Multiple stellar populations (MSPs) have been discovered in many globular clusters (GCs) \citep[e.g.][]{Carretta2009a,Piotto2015}.
Due to the high precision HST UV Legacy Survey of Galactic Globular Clusters, the knowledge of the MSP phenomenon has been significantly enriched \citep[e.g.][]{Piotto2015}.
The MSPs challenge the traditional understanding that all stars in a GC were born in the same environment and the complex chemical element features of the MSP raises puzzles about their formation scenario.
\cite{Renzini2015} and \cite{Bastian2018} provide recent reviews of the observational features of MSPs and find that currently MSP models have difficulty accounting for the observational data.

The Na-O, N-C and Al-Mg (sometimes) anti-correlations are the key feature of the MSP phenomenon, with Na-N and C-O being positively correlated.
The second population stars have enhanced N, Na and Al with depleted C, O and Mg compared to the first population or field-like stars.
On the other hand, the sum C+N+O remains constant \citep[e.g.][]{Dickens1991} and the spread of Fe is also narrow in most GCs \citep[e.g.][]{Carretta2009b,Bailin2019}.
The Helium (He) abundance also varies in GCs and correlates with Na and Al \citep[e.g.][]{Villanova2009,Dupree2013,Gratton2013,Mucciarelli2014}.
However, the degree of the element spread has a large variation in different GCs \citep[e.g.][]{Carretta2009a} suggesting that a number of processes play a role which may be variably active depending on the detailed physcial conditions at hand.

The measurement of age differences between different populations have a large uncertainty.
Based on HST photometry, \cite{Nardiello2015} estimated the age spread of MSPs in NGC~6352 to be $10\pm120$~Myr and the upper limit is about $300$~Myr.
The same limit is also found in NGC~6656 \citep{Marino2012}.

In many GCs the second populations are more concentrated radially compared to the first populations \citep[e.g.][]{Sollima2007,Lardo2011}.
However, some GCs show no radial difference of their MSPs \citep[e.g.][]{Milone2009}.
A few exceptions even show an opposite trend, namely that the first populations are more concentrated \citep[e.g.][]{Larsen2015}.

MSPs can be identified from the color spread in the color-magnitude diagram of a GC \citep[CMDs; e.g.][]{Gratton2004,Gratton2012,Han2009,Milone2017,Niederhofer2017a,Niederhofer2017b,Martocchia2018b,Bonatto2019} and the chromosome diagram \citep[e.g.][]{Milone2015,Milone2017,Milone2018,Milone2019,Lagioia2019}.
Using the chromosome diagram, sub-populations are found and identified as discrete groups \citep[e.g. NGC~2808,][]{Milone2015}.

The properties of MSPs weakly depends on the luminosity function and the locations of their host GCs in the Galaxy \citep{Carretta2010}.
But the MSP phenomenon strongly depends on the properties of star clusters.
HST photometry of 57 GCs indicates that the properties of MSPs strongly correlate with the present-day total mass of a GC, as massive GCs tend to have a larger He, N spread and a larger fraction of second population stars \citep{Milone2017, Milone2018}.
These phenomenons suggests that the formation of MSPs is more related to the self-evolution of GCs and is less influenced by the galactic environment. 
The Gaia-ESO survey iDR4 data suggests that the Mg-Al anti-correlation is not seen in all GCs and disappears for the less massive or most metal-rich GCs \citep{Pancino2017}.
On the other hand, MSPs appear in old massive GCs but not metal-rich and young massive star clusters.
The occurrence of MSPs strongly depends on the age of star clusters and on its iron abundance, [Fe/H] \citep{Krause2016,Bastian2018}.
Young massive star clusters with ages $<2$~Gyr \citep[e.g][]{Martocchia2018a,Martocchia2018b,Martocchia2019} have not been found to have MSPs.
However, the exact age boundary is still unclear.
A recent work of \cite{Li2019} find that the Lindsay 113, a $4$~Gyr old cluster, has no evidence of MSPs. 
Metal-rich low-mass (open) clusters with [Fe/H]$>-0.5$ and mass below $2-5\times10^{4}$~M$_\odot$ also do not show the feature of MSPs \citep[e.g.][]{Bragaglia2012,Bragaglia2014,Bragaglia2018}.

The observed narrow spread of Fe suggests that the MSPs are neither due to mergers of two star clusters with different Fe abundance nor a result of the accretion of surrounding inter-stellar medium \citep[e.g.][]{Pflamm2009}.
The radially concentrated phenomenon of second populations suggests that they were likely formed in the self-enriched gas at the center of clusters, where the element abundance is polluted by the stellar winds of the first populations (polluters).
Since supernovae generate Fe, the second populations forms either in the primordial gas before supernovae (during the first $5$~Myr after or during the first population forms in a GC) or in the retained (low-velocity) stellar winds of polluters after the primordial gas is cleared away by supernovae.
Depending on models, the first/second populations can form simultaneously during a continuing star formation or form one after another.
Thus to avoid confusion, hereafter we distinguish MSPs as primordial/enriched populations instead of first/second populations.
There are four major potential polluters that have been frequently discussed so far:

a) Intermediate-mass asymptotic giant branch (AGB) stars \citep[e.g.][]{Cottrell1981,DErcole2010,DAntona2016} are considered as such potential polluters.
After about 30~Myr, stars from the primordial population evolve through the AGB phase.
The slow stellar winds of AGB stars, which are element enriched, can accumulate in the GC potential center and form new stars.
One issue of the AGB scenario is that they do not predict the Na-O anti-correlation, but a Na-O correlation.
Hence, a GC needs to re-accrete unmixed gas after supernova explosions and it is unclear where this gas comes from.
The mass range in the AGB scenario is about $6 - 8$ M$_\odot$.
Stars of lower mass do not contribute, because they do not undergo hot bottom burning.
Thus, another major issue of the AGB scenario is that the enriched populations would only constitute $2-10\%$ of the total mass of a GC, which is much lower than the observational fractions \citep[more than half;][]{Milone2017}.
This is called the ``mass budget problem''.
This problem also appears in other scenarios where enriched populations are assumed to form from the low-velocity stellar winds generated by polluters.
If the AGB scenario contributes, either the IMF of the primordial population stars needs to be very top-heavy \citep[e.g.][]{Prantzos2006,Bekki2017}, and$/$or a significant amount of FG stars are lost \citep[e.g.][]{Decressin2008}.

b) Fast rotating massive main sequence stars (FRMS; $>20$~M$_\odot$) are considered as one possible polluter \citep{Decressin2007}.
The massive stars can have a high enough temperature to trigger the H-burning with Ne-Na and Mg-Al chains.
The rotational mixing can carry out the elements from the convective core, thus their stellar winds can have a Na-O and N-C anti-correlation.
On the other hand, the wind velocity of FRMS can be low enough to be captured by the gravitational potential of GCs.
Thus enriched populations can form from these elementally enriched winds, similar as the AGB scenario.
This mechanism has a problem to explain the Al-Mg anti-correlation, unless the depletion process via proton capture on $^{24}$Mg is $1000$ times enlarged.

c) \cite{deMink2009} suggested that interacting massive binaries may also contribute to element enrichment (here-after referred as the BINARY channel).
The mass transfer between the two companion stars results in the low-velocity ejection of the envelope of the primary star, which contains the enriched elements with Na-O anti-correlation.

d) \cite{Denissenkov2014} introduced a super massive main sequence (SMS) star as another type of polluter.
The idea is that in the early stage of a dense GC, massive OB stars can suffer runaway collisions, which leads to the formation of SMS stars ($>10^3$~M$_{\odot}$) at the center of clusters \citep{PZ2004}.
One advantage of this scenario is that for a mass range of $10^3-10^4$~M$_{\odot}$, the temperature reaches $7 \times 10^7 - 8 \times 10^7$~K for H-buring to produce a consistent Al-Mg anti-correlation \citep{Prantzos2017}.
Following up on this idea, \cite{Gieles2018} suggested that if  continuous gas accretion and collisions are considered during the embedded phase, the SMS can provide an order of magnitude higher polluted gas mass than its stellar mass.
This can help to solve the mass budget problem.

\cite{Bastian2015} investigate the He abundance and Na-O anti-correlation of these four scenarios and compare with the observations and found that the individual scenarios fail to provide a consistent abundance of He, Na and O.
Especially, in order to fit the observed Na-O data, He is overproduced in the AGB, FRMS and SMS scenarios.

Although each model has difficulty to explain all the observational phenomenon of MSPs, it seems natural that several scenarios can work together.
Most of the MSP formation models suggested in previous works discuss the detailed element evolution of specific objects only, but do not self-consistently perform dynamical simulations of GCs.
It is important to also consider the dynamical process in young GCs when discussing the MSP formation scenario.
For example, in the AGB scenario, to solve the mass budget problem, a very top-heavy IMF is suggested to generate enough AGB stars \citep[e.g][]{Prantzos2006,Bekki2017}.
This is a problematical proposition for GCs because the strong stellar winds of massive stars lead to significant mass loss and the GCs are unlikely to survive for a Hubble time due to the black hole (BH) subsystem heating the stellar cluster \citep[e.g][]{Fukushige1995,Mackey2008,Chatterjee2017,Giersz2019,Wang2019}.
This possibility is usually not considered carefully in the MSP formation scenario, which is discussed without the dynamical modelling of GCs.

On the other hand, the BINARY and SMS scenarios are very sensitive to the stellar dynamical processes, especially through the mergers of binary stars.
Since the binary orbit contains a large angular momentum, a binary merger is likely to produce a FRMS \citep{deMink2013}.
Thus, stellar mergers/collisions naturally link these scenarios (BINARY, SMS and FRMS) together and can be investigated via $N$-body simulations.

The observations of young star formation regions indicate a large fraction of stellar multiplicities in dynamically young populations \citep[][and references there in]{Duchene2013}.
If GCs had a similar multiplicity property when they formed, they should have contained a large fraction of primordial binaries initially \citep{Kroupa1995a,Leigh2015,Belloni2017}.
On the other hand, during the long-term evolution of GCs, the cluster will have expanded due to the initial gas expulsion \citep[e.g.][]{Baumgardt2008b}, the stellar-wind mass loss and the dynamical heating of cluster cores, especially from BH subsystems \citep[e.g.][]{Breen2013,Mackey2008,Chatterjee2017,Giersz2019}.
Thus the present-day density of GCs is expected to be much smaller compared to the initial state.
Therefore, we expect initially GCs to have been compact with a large fraction of binaries.
In such a dense environment, the dynamical encounters between binaries and stars can result in significant changes of the binary orbits and in binary mergers \citep[e.g.][]{Banerjee2012b,Oh2018}.
Especially, during the first $5$~Myr when the clusters are still in the embedded phase, it is possible that new stars formed in the enriched gas polluted by the mergers, i.e., the ejection of enriched stellar material during stellar mergers (BINARY) and the stellar winds of new stars (FRMS and SMS) after mergers.
Considering a typical star formation efficiency (SFE) of about $0.1-0.3$ \citep[e.g.][]{Lada2003,Megeath2016} for the primordial population, enough gas remains to form a large number of element-enriched enriched populations.
This may provide a solution to the mass budget problem.

After the gas expulsion due to the energetic supernova starting around $5$~Myr, the cluster loses potential energy and its density decreases which reduces the merger rate.
Without the primordial gas, enriched populations can only continue to form from the slow winds of stars (including the AGB scenario), which only contribute a small fraction of MSPs.
Thus in our multi-channel scenario, we assume that the major fraction of the enriched populations forms during the embedded phase.
In this sense, MSPs may not be truly distinct populations, but they might appear in an overall continuous star formation process (except for the stars formed later during the AGB phase), i.e, enriched stars form from more element-enriched gas polluted by the ejection/winds from the binary mergers of massive stars.

In this work, the dynamical modelling of massive star clusters with a high fraction of primordial binary population is carried out to help for clarifying the contributions of the three MSP formation processes (BINARY, SMS and FRMS) before the gas expulsion after supernova.
In Section~\ref{sec:method}, we describe the method and the initial conditions.
The analysis of the merger properties is explained in Section~\ref{sec:mergers}.
In Section~\ref{sec:polluters}, the MSP models are discussed based on our results.
In Section~\ref{sec:discussion}, we discuss the impact of our assumptions applied in the simulations.
Finally, Section~\ref{sec:conclusion} contains the conclusion.

\section{Methods}
\label{sec:method}

\subsection{N-body models}

\subsubsection{Direct $N$-body code}

In this work, stellar collisions and coalescences driven by few-body interactions are the key process to produce the element enrichment.
Thus we perform the computations by using the direct $N$-body code \textsc{nbody6++gpu} \citep{Wang2015}, in which the regularization algorithms \citep{KS1965,Mikkola1999} are implemented to ensure an acceptable accuracy in the treatment of few-body interactions.
\textsc{nbody6++gpu} is a MPI parallelization optimized version of the state-of-the-art code \textsc{nbody6} \citep{Aarseth2003,Nitadori2012}.
This series of codes include the single and binary stellar evolution packages (\textsc{sse/bse}) from \cite{Hurley2000,Hurley2002} \footnote{We used the original \textsc{sse/bse} version (before 2016) in this work and did not include the update of SSE/BSE implemented in the \textsc{nbody6} code.}.

In the \textsc{nbody6++gpu} code, there are two definitions of a merger: stellar collision and stellar coalescence.
The stellar collision is defined as a merger without a common envelope stage.
Usually this type of merger has a high-eccentric or hyperbolic orbit.
The stellar coalescence is a merger with a common envelope evolution.
The binary orbit in this case is usually circularized before the merger.

During the merging process, a certain amount of mass is lost.
Several previous studies based on hydrodynamic simulations of mergers have investigated the mass loss through binary collisions for blue straggler formation \citep[e.g.][]{Lombardi2002,Glebbeek2013}.
However, for massive stars above $100$~M$_\odot$, how the merger evolves is not well understood.
\cite{deMink2014} assumed that to remove excess angular momentum in order to make the newly formed star stable, roughly $0-25\%$ (typical $10\%$) of the binary mass is lost during the coalescence.
In the case of the MS-MS stellar collision implemented in \textsc{nbody6(++)}, if the binary orbital kinetic energy is larger than the internal binding energy of the new formed star and the peri-center distance is less than the averaged stellar radius of the two progenitors (binary components), the mass loss is fixed to be $30\%$ of the secondary mass of a binary, 
otherwise no mass loss is assumed\footnote{Due to a small bug in the version of \textsc{nbody6++gpu} used in this work, some high-eccentric low-mass binary mergers miss the $30\%$ mass loss. This is identified at the end of all finished simulations, thus difficult to fix. Fortunately this does not have an influence on our results due to the small number of cases. On the other hand, the mass loss from the merged stars is very uncertain and the assumption of $30\%$ is merely used here as a reasonable value to allow an assessment of the mass accumulating in a young GC as a consequence of stellar mergers}.
The mass loss during a stellar coalescence (common envelope and Roche mass-transfer) is based on the implementation in the \textsc{bse} package.
Depending on the mass ratio of the two companion stars, the maximum mass loss rate is $15\%$ of the total mass of the binary.

\subsubsection{Initial conditions}
\label{sec:init}

The initial conditions of GCs are uncertain.
Due to the $N$-body modelling of GCs being very CPU expensive \citep{Wang2016}, a large series of simulations to cover all parameter space is not possible.
With limited computational resources, the best we can do is try to apply constraints on the initial conditions based on observational data from young embedded star clusters combined with observational knowledge of GCs.
We follow the configuration of initial conditions from \cite{Wang2019}, where Orion Nebula Cluster (ONC) like young star clusters are studied, applying their observational constraints on the initial models.
The notion here is that while GCs are born significantly more massive than the ONC or even R136 in the LMC \citep{Banerjee2012}, the basic physics and processes remain the same.

The stellar merging rate in young GCs depends on the properties of the initial multiplicity (binaries and high-order systems).
With the assumption that the primordial binary property is universal, \cite{Kroupa1995a,Kroupa1995b,Kroupa2011} used the method of inverse dynamical population synthesis to construct the initial binary period and mass-ratio distribution in young star clusters.
The universal primordial binary property is also supported by the observational data of the binary properties in young star clusters \citep{Duchene2018} and by the analysis of the present-day binary population in observed GCs \citep{Leigh2015,Belloni2017}.
For the initial distribution of high-mass (OB star) binaries ($>5$~M$_\odot$), the observational constraints from \cite{Sana2012} are used.
The details of the binary mass ratio, eccentricity and period distribution are summarized in \cite{Kroupa2013,Belloni2017}, hereafter referred to ``KSB''.

One important feature of this binary model is that the periods for both low-mass and high-mass binaries can reach less than one day.
If the period and eccentricity distributions are independent of each other, some of the binaries would have the two stars touching each other at the peri-center position and should merge within one orbit. 
\cite{Kroupa1995b} suggested pre-main-sequence eigenevolution to redistribute the energy and angular momentum of such binaries.
Thus most of these binaries would not be in the forbidden period-eccentricity region.
Following this idea, \cite{Belloni2017} derived an upgraded version of pre-main-sequence eigenevolution and underlined the universality hypothesis based on observational constraints from the stellar color distribution in GCs.
However, in the $N$-body simulations with the \textsc{bse} package, some binaries with small peri-center distances can still immediately merge at the zero-age main sequence (ZAMS) stage. 
Thus they should already merge at the pre-main-sequence stage.
These mergers probably cannot be distinguished from normal single stars (without considering the stellar rotation).
Thus, in our analysis, we distinguish these pre-main-sequence stellar-evolution (PSE) mergers from other cases.

Following \cite{Wang2019}, we assume the star formation process is highly self-regulated, i.e., individual stars determine their mass growth based on the local environment (gas clouds) and the action of stellar winds and outflows \citep{Adams1996}.
This gives strong constraints on the IMF of stars within clusters, the initial cluster half-mass radius ($R_{\mathrm{h,0}}$) and the initial mass distribution of the star clusters.
The $R_{\mathrm{h,0}}$ is determined by the $M_{\mathrm{ecl}}$ - $R_{\mathrm{h,0}}$ relation derived by \cite{Marks2012}, who use the observed binary star energy distribution to constrain the highest density that star clusters could have had:
\begin{equation}
  \label{eq:rhmecl}
  R_{\mathrm{h,0}}[pc] = 0.1^{+0.07}_{\mathrm{-0.04}} (M_{\mathrm{ecl}}[M_{\odot}])^{0.13\pm0.04},
\end{equation}
where $M_{\mathrm{ecl}}$ is the total initial stellar mass.
The canonical initial mass function (IMF) from \cite{Kroupa2001} with optimal sampling \citep{Kroupa2013,Yan2017} is applied.
The ALMA observation of Serpens South, a young star-forming region, shows a high-degree of mass segregation \citep{Plunkett2018}.
$N$-body models suggest that this feature may be more consistent with primordial fully-mass segregation \citep{Pavlik2019}.
Thus, our initial models are fully-mass-segregated using the method from \cite{Baumgardt2008}.

We apply the spherically symmetric Plummer model to generate the distribution of the positions and velocities of the stars.
A Solar-neighborhood tidal field is applied to determine the tidal radii of the clusters.

The large binary fraction (close to unity) based on the models of \cite{Kroupa1995a,Kroupa1995b,Belloni2017} makes the $N$-body simulations significantly challenging.
The current \textsc{nbody6} codes cannot deal with a very large fraction of binaries (when $N>10^{5}$ stars) due to a lack of an efficient parallelization method for binary dynamics.
Thus realistic models of GCs with $10^6$ stars are very time-consuming and accuracy limited.
Although we only consider the first $5$~Myr life of the GCs \citep[before the first supernova and thus gas expulsion; ][]{Baumgardt2008b}, the computations are still very difficult.
Due to this limit, we perform a set of models with different $M_{\mathrm{ecl}}$ (lower than typical massive GCs) and extrapolate the results to the cases of massive GCs to obtain an informed guess of the likely role of stellar mergers.
For each low $M_{\mathrm{ecl}}$, we perform several models to obtain better statistics.

We also compute a few models with a high $M_{\mathrm{ecl}}$ without any primordial binaries for comparison.
The major parameters of the initial conditions are listed in Table~\ref{tab:sets}.
The initial relaxation timescale, $T_{\mathrm rh,0}$, at the initial half-mass radius, $R_{\mathrm rh,0}$, is calculated by using the \cite{Spitzer1987} formula 
\begin{equation}
  T_{\mathrm rh} = 0.138 \frac{N^{1/2} R_{\mathrm h}^{3/2}}{\langle m \rangle^{1/2} G^{1/2} \ln \Lambda},
\end{equation}
with the Coulomb logarithm being $\ln \Lambda=12$ and an averaged initial stellar mass, $\langle m \rangle=M_{\mathrm{ecl}}/N_{\mathrm{0}}$.
Since the binary period distribution has a very wide range, many long-period binaries are dynamically equal to two single stars.
Thus we estimated two values for $T_{\mathrm rh,0}$, with resolved and unresolved binaries.
In the resolved case, the initial total number of stars, $N_{\mathrm 0}$, treats one binary as two stars (the real number of stars in the system).
In the unresolved case, a binary is treated as one single (center-of-mass) star.

\begin{table*}
  \centering
  \caption{Initial parameters of the $N$-body models. The low-mass models are computed $N_{\mathrm model}$ times for statistical analysis. $M_{ecl}$ is the initial total stellar mass. $N_{\mathrm 0}$ is the initial total number of stars. $R_{\mathrm rh,0}$ is the initial half-mass radius. $T_{\mathrm rh,0}$ is the initial two-body relaxation timescale at $R_{\mathrm rh,0}$. In the unresolved cases, binaries are treated as center-of-mass stars. $m_{max,0}$ is the initial maximum stellar mass sampled from the IMF according to optimal sampling. The ``B'' and ``S'' in the model name indicates that the model contains initially $100\%$ binary or pure single stars. The ``M'' indicate an initially mass-segregated model.
  }
  \label{tab:sets}
  \begin{tabular}{@{}llllllllll@{}}
    \hline
    Model & B-M5K & B-M10K & B-M20K & B-M40K & B-M80K & B-M160K & S-M80K & S-M160K & S-M320K \\
    \hline
    $N_{\mathrm Model}$ & 11 & 5 & 3 & 1 & 1 & 1 & 1 & 1 & 1\\
    $M_{\mathrm ecl}[\mathrm{M}_\odot]$ & 5000 & 10000 & 20000 & 40000 & 80000 & 160000 & 80000 & 160000 & 320000 \\
    $N_{\mathrm 0}$ & 8763 & 17329 & 34426 & 68587 & 136879 & 273427 & 136879 & 273427 & 546476 \\
    $R_{\mathrm h,0}[\mathrm{pc}]$ & 0.30 & 0.33 & 0.36 & 0.40 & 0.44 & 0.48 & 0.43 & 0.47 & 0.52 \\
    $T_{\mathrm rh,0}[\mathrm{Myr}]$ (resolved) & 3.45 & 5.54 & 8.89 & 14.3 & 23.2 & 37.5 & 23.5 & 38.1 & 61.6 \\
    $T_{\mathrm rh,0}[\mathrm{Myr}]$ (unresolved) & 2.46 & 3.94 & 6.31 & 10.1 & 16.4 & 26.5 & 23.5 & 38.1 & 61.6 \\
    $m_{max,0}[\mathrm{M}_\odot]$ & 90 & 110 & 126 & 136 & 142 & 146 & 142 & 146 & 148 \\
    Binary & \multicolumn{6}{l}{$100\%$ primordial binary (KSB)} & \multicolumn{3}{c}{No binary}\\
    Profile & \multicolumn{9}{l}{Plummer models}\\
    IMF & \multicolumn{9}{l}{canonical IMF with optimal sampling}\\
    Tidal field & \multicolumn{9}{l}{solar-neighborhood}\\
    Mass-segregation & \multicolumn{9}{l}{initially fully mass-segregated} \\
    \hline
  \end{tabular}
\end{table*}

The metallicity also influences the binary stellar evolution and mass loss of stars.
However, our knowledge of the stellar evolution of massive stars and binaries is very uncertain and the current \textsc{bse} package has limitations in providing proper evolution tracks for low-metallicity stellar winds, stellar rotation and the evolution of massive ($>150$~M$_\odot$) stars.
Thus we only investigate a metallicity of $Z \approx 0.001$, which represents the case of a typical GC.
The high computational cost is another reason to choose only one metallicity in this work.

\section{Mergers}
\label{sec:mergers}

\subsection{Two merger types}

\begin{figure}
  \includegraphics[width=1.0\columnwidth]{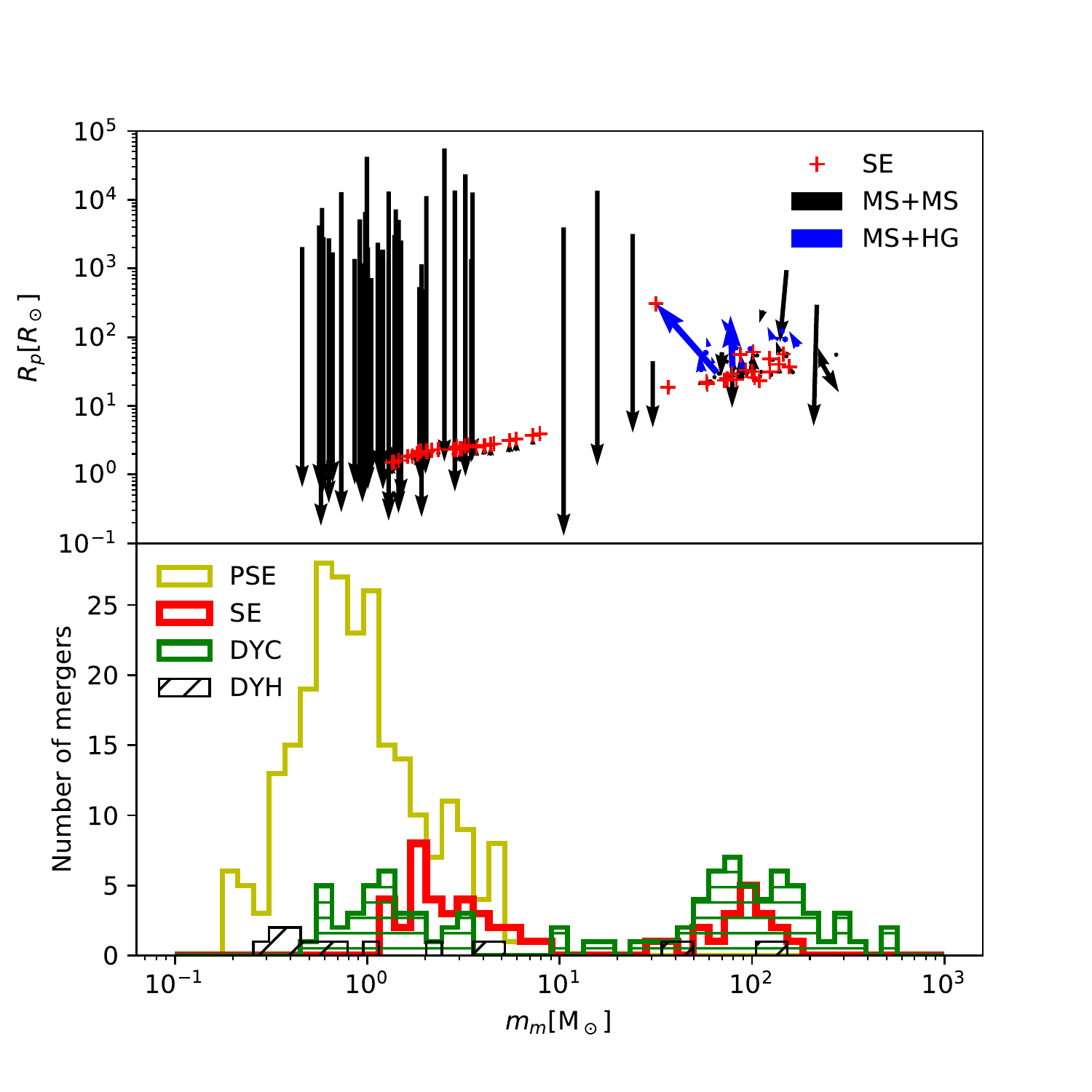}
  \caption{The diagram shows the two types of mergers in the model B-M160K.
    Upper panel: the peri-center distance $R_{\mathrm p}$ of merging binaries at the age of $0$~Myr and at the time before the merging.
    The arrows indicate the direction of evolution of $R_{\mathrm p}$.
    The x-axis shows the total mass of the binary before the merger.
    The label ``SE'' indicates the mergers dominated by binary stellar evolution.
    The black lines (dots) indicate the mergers of two MS stars ``MS+MS'' and the blue lines indicate the mergers of a MS star and a Hertzsprung-gap (HG) star.
    Both blue and black colours refer to DY mergers.
    Lower panel: the number of mergers due to binary stellar evolution (SE), the dynamically induced mergers with closed orbits (DYC) and the dynamically induced mergers with hyperbolic orbits (DYH).
  }
  \label{fig:types}
\end{figure}

In models with primordial binaries (``B-'' series), we identify two types of mergers: the stellar-evolution driven and the dynamically driven types.
As mentioned in Section~\ref{sec:init}, there are also PSE mergers at time zero.
Even if we exclude these, some binaries can still merge in a short timescale (in a few Myr) without any significant influence through star-cluster dynamical perturbations.
We name these mergers as stellar-evolution-driven (SE) mergers.
By using the stand-alone \textsc{bse} code \citep{Hurley2002}, we can identify all PSE and SE mergers from the initial conditions.
The stellar encounters between binaries and singles can significantly perturb longer period binary orbits and induce dynamically driven (DY) mergers, including the binary mergers with closed Kepler orbits (DYC) and hyperbolic mergers with open orbits (DYH).
Here we list all types of mergers with their definition:
\begin{itemize}
\item PSE: stellar-evolution-driven merger before the zero-age-main-sequence,
\item SE: stellar-evolution-driven merger during or after the zero-age-main-sequence,
\item DY: dynamically driven merger,
\item DYC: dynamically driven merger on a closed Kepler orbit,
\item DYH: dynamically driven merger on a hyperbolic orbit.
\end{itemize}
The SE and DY mergers can be easily distinguished by the evolution of the binary peri-center distance, $R_{\mathrm p}$, from the beginning (0~Myr) to the time before the merger.
Fig.~\ref{fig:types} shows the $R_{\mathrm p}$ changes of SE and DYC mergers depending on the total mass of the binary, $m_{\mathrm 12}=m_{\mathrm 1}+m_{\mathrm 2}$ (before merging) for the most massive computed cluster model with primordial binaries (B-M160K).
$m_{\mathrm 1}$ and $m_{\mathrm 2}$ are the primary and secondary masses respectively.
It can be seen that for the low-mass binaries ($m_{\mathrm 12}<10~M_\odot$), the SE mergers have negligible changes of $R_{\mathrm p}$, while the DY mergers have significant changes.
Thus most of the SE mergers suffer from a weak influence from dynamical perturbations.
The SE mergers are also well aligned, indicating a merging boundary (a minimum $R_{\mathrm p}$) for different $m_{\mathrm 12}$ values.

In the lower panel of Fig.~\ref{fig:types}, the number of mergers for different merging channels and different $m_{\mathrm 12}$ is shown.
There is a clear signal of a bimodal distribution of low-mass and high-mass binaries separated by $m_{\mathrm 12}= 10 M_\odot$.
This is due to the different initial primordial-binary properties of low-mass stars \citep[$m<5~M_\odot$; based on ][]{Kroupa1995a,Kroupa1995b} and high-mass stars \citep[$m>5~M_\odot$; based on ][]{Sana2012}.
The PSE mergers dominate the low-mass region.
Also, the SE mergers contribute a significant fraction in both low-mass and high-mass regions.
In the high-mass region, the number of DYC mergers is larger.
Only few DYH mergers are distributed over the whole mass range.

\begin{figure}
  \includegraphics[width=1.0\columnwidth]{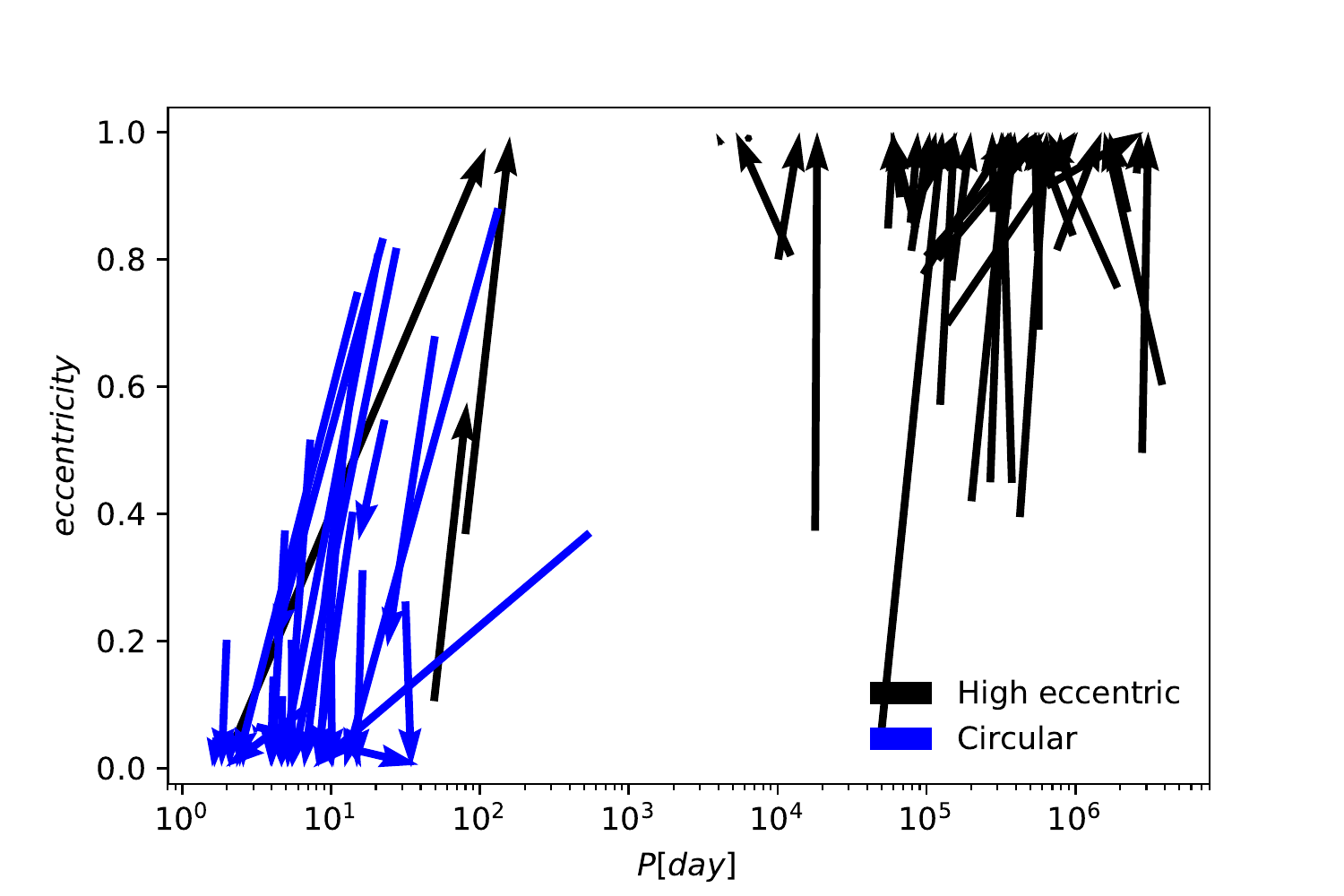}
  \caption{The evolution of eccentricities and periods of the ``DYC'' type of mergers in the model B-M160K.
    The black lines show the high-eccentric binary mergers (produced after encounters increase the eccentricity) and the blue lines show the circularized binary mergers (produced after encounters drive a binary onto a tidal-circularization phase and then merger with common-envelope and mass-transfer evolution).
  }
  \label{fig:pecc}
\end{figure}

For the DYC mergers, there are two evolution trends separated by the circularization process.
In the eccentricity-period evolution diagram (Fig.~\ref{fig:pecc}), a clear bimodal eccentricity evolution can be identified together with the period evolution.
The short-period binaries tend to have circularized orbits before merging and the long-period mergers all have high pre-merger orbital eccentricity.

\subsection{Mergers -- star cluster mass}
\label{sec:nmecl}

\begin{figure}
  \includegraphics[width=1.0\columnwidth]{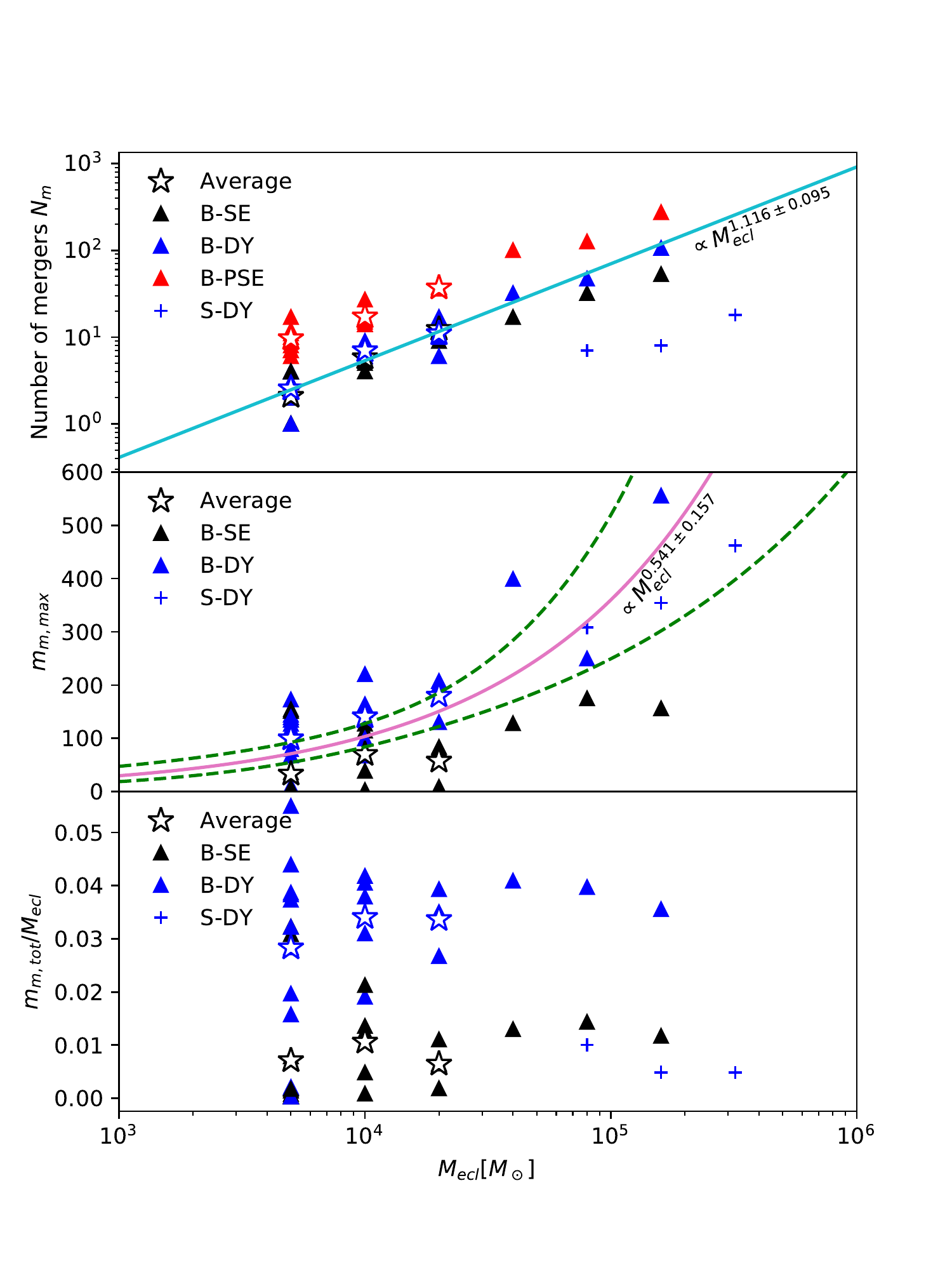}
  \caption{The merger properties depending on the initial stellar mass of the clusters, $M_{\mathrm{ecl}}$.
    Upper panel: the number of mergers through different channels. The stars with the respective colors indicate the averaged number for models computed more than once ($N_{\mathrm{Model}}>1$, Table~\ref{tab:sets}).
    The prefixes ``B-'' and ``S-'' indicate the model with and without primordial binaries respectively.
    The lightblue line is the fitting curve (Eq.~\ref{eq:fit}) of the number of mergers for the type B-DY.
    Middle panel: the maximum mass of the merged stars, $m_{\mathrm{m,max}}$ (by summation of the two components' masses of the progenitor binary).
    The purple line is the fitting curve of $m_{\mathrm{m,max}}$ for both B-DY and S-DY data combined (Eq.~\ref{eq:fit}).
    The dashed lines show the $1\sigma$ boundary based on the fitting curve.
    Lower panel: the ratio between total stellar masses involved in mergers, $m_{\mathrm{m,tot}}$, and $M_{\mathrm{ecl}}$.
  }
  \label{fig:nmecl}
\end{figure}

In Fig.~\ref{fig:nmecl}, we analyze the merger properties in dependence of $M_{\mathrm{ecl}}$ for our models.
The upper panel shows the number of mergers, $N_{\mathrm m}$, from different channels.
In the $\ln N_{\mathrm m}$-$\ln M_{\mathrm{ecl}}$ space, there is a clear linear dependency for all merger types with similar slopes and different offsets along the y-axis ($N_{\mathrm m}$).
The PSE mergers dominate, similarly as already seen in Fig.~\ref{fig:types}.
The DY mergers are in the middle.
In the models without primordial binaries, the DY mergers are the only channel.
It is obvious that the existence of a large fraction of primordial binaries significantly enhances the number of mergers for the same $M_{\mathrm{ecl}}$.
We apply a non-linear least-squares fitting method \footnote{We use the \textsc{python} module \textsc{scipy.optimize.curve\_fit} to do the fitting.} using the function
\begin{equation}
  \label{eq:fit}
  \ln N_{\mathrm{DY}} = \alpha \ln M_{\mathrm{ecl}}[M_\odot] + \beta
\end{equation}
to the data of the DY type mergers, $N_{\mathrm{DY}}$, for models with primordial binaries (B-DY).
The fitting coefficient, $\alpha(N_{\mathrm{DY}}) = 1.116 \pm 0.095$, indicates a linear dependence.
Since other types of mergers have similar slopes as the case of $N_{\mathrm{DY}}$, the number of mergers, $N_{\mathrm m}$, (all different types) almost depends linearly on $M_{\mathrm{ecl}}$.
For the PSE and the SE mergers, this is reasonable because the number of binaries depends linearly on $M_{\mathrm{ecl}}$ and it is expected to have the same dependence of mergers due to that the initial binary population is universal.
For the DY mergers, the dynamical perturbation can change the binary orbit in or out of the orbits for mergers.
It is expected that the perturbations correlate with the two-body relaxation process.
In clusters with different $M_{\mathrm{ecl}}$, the relaxation timescale is different.
Thus the ratio $N_{DY}/M_{\mathrm{ecl}}$ is expected to increase with decreasing relaxation timescale with a $M_{\mathrm{ecl}}$ dependence.
The almost linear dependence indicates rather that $N_{DY}$ is more sensitive to the number of binary sources instead of the perturbation processes.
We can fix $\alpha=1.0$ and fit $N_{\mathrm{DY}}$ again.
The coefficient, $\beta=-7.528 \pm 0.093$, for the DY type mergers.
Thus we obtain
\begin{equation}
  \label{eq:ndy}
  \begin{split}
    N_{\mathrm{DY}}& =  e^{\beta} M_{\mathrm{ecl}}[M_\odot] \\
                 & =  (5.379 \pm 0.500)\times 10^{-5} M_{\mathrm{ecl}}[M_\odot] .
  \end{split}
\end{equation}

SMS stars can form via mergers.
In the middle panel of Fig.~\ref{fig:nmecl}, the dependence of the maximum merger mass, $m_{\mathrm{m,max}}$, on $M_{\mathrm{ecl}}$ is shown.
It is obvious that the most massive mergers all come from the DY type and they are generated by multiple mergers.
The maximum mass of mergers from the PSE/SE channels is limited by the maximum mass allowed for the two component stars of the primordial binaries.
The optimal sampling method of the IMF \citep{Kroupa2013,Yan2017} results in a maximum stellar mass of $150$~M$_\odot$.
Thus the PSE/SE channel cannot produce mergers with $m_{\mathrm{m,max}}>300$~M$_\odot$, and the mergers with masses more than this limit can only form via multiple mergers through the DY channel.
In the fully mass-segregated models used here, the most massive stars are initially in the star cluster center.
Thus the multiple mergers are expected in a highly dense region, which is consistent with the previous $N$-body modelling of dense star clusters \citep{PZ2004}.

There is a significant scatter of $m_{\mathrm{m,max}}$ (middle panel in Fig.~\ref{fig:nmecl}).
This high uncertainty in $m_{\mathrm{m,max}}$ at a given $M_{\mathrm{ecl}}$ is understood as the few-body dynamics can result in mergers but also in ejections.
If the massive binaries were ejected out of the center of the star cluster, $m_{\mathrm{m,max}}$ would be small, like in the case of B-M80K.
Although the uncertainty is rather large, there is still a trend that more massive star clusters produce larger $m_{\mathrm{m,max}}$ values.
Interestingly, this trend seems independent of the primordial binary properties, as both the S- (models without primordial binaries) and B- models with the same $M_{\mathrm{ecl}}$ produce consistent $m_{\mathrm{m,max}}$.

The purple curve in the middle panel of Fig.~\ref{fig:nmecl} is obtained by fitting Eq.~\ref{eq:fit} after replacing $N_{\mathrm{DY}}$ by $m_{\mathrm{m,max}}$ for all B-DY and S-DY data.
The fitting formula with mean parameters is
\begin{equation}
  \label{eq:mmax}
  \ln m_{\mathrm{m,max}}[M_\odot] = 0.541 \ln M_{\mathrm{ecl}}[M_\odot] - 0.341 .
\end{equation}
The fitting parameters, $\alpha(m_{\mathrm{m,max}}) = 0.541 \pm 0.157$ and $\beta(m_{\mathrm{m,max}}) = -0.341 \pm 1.520$, have quite a large uncertainty.
By using the fitting result of the convariance matrix,
\begin{equation}
  \label{eq:cov}
  \begin{bmatrix}
    \sigma_{\mathrm{\alpha \alpha}}^2 & \sigma_{\mathrm{\alpha \beta}}^2 \\
    \sigma_{\mathrm{\beta \alpha}}^2 & \sigma_{\mathrm{\beta \beta}}^2 \\
  \end{bmatrix}
  (m_{\mathrm{m,max}})  =
  \begin{bmatrix}
    0.0248 & -0.237 \\
    -0.237 & 2.31 \\
  \end{bmatrix},
\end{equation}
we can construct the $1~\sigma$ region of the fitting curves (shown as the region between the two dashed lines in Fig.~\ref{fig:nmecl}).
Here $\sigma_{\mathrm{\alpha \alpha}}$ and $\sigma_{\mathrm{\beta \beta}}$ are the one standard deviation errors of $\alpha$ and $\beta$ and $\sigma_{\mathrm{\alpha \beta}}$ or $\sigma_{\mathrm{\beta \alpha}}$ represents the correlation between $\alpha$ and $\beta$ respectively.
The convariance matrix defines the $1~\sigma$ boundary of $\alpha$ and $\beta$ in the $\alpha$-$\beta$ parameter space (an ellipse).
Thus, the maximum and minimum $m_{\mathrm{m,max}}$ depending on $M_{\mathrm{ecl}}$ inside this boundary can be evaluated.
The higher $M_{\mathrm{ecl}}$ end has a wider spread due to the lack of data points and the large scatter of $m_{\mathrm{m,max}}$.
Although the spread is rather large, the possible $m_{\mathrm{m,max}}$ values for massive GCs can be estimated.
For $M_{\mathrm{ecl}}=10^{6} M_{\odot}$, $m_{\mathrm{m,max}} \approx 1250_{-628}^{+1263} M_{\odot}$.
Thus, with this extrapolation, a SMS star MSP polluter with about $10^3 M_{\odot}$ can possibly form in a massive young GC.
The more massive the GC, the more massive the SMS is expected to form.

In the lower panel of Fig.~\ref{fig:nmecl}, we show the ratio of the total mass of stars involved in all mergers within the first $5$~Myr, $m_{\mathrm{m,tot}}$, in dependence of $M_{\mathrm{ecl}}$.
There is also a large scatter of $m_{\mathrm{m,tot}}$ for the same $M_{\mathrm{ecl}}$, which is mainly contributed by the scatter of the massive mergers.
The $m_{\mathrm{m,tot}}/M_{\mathrm{ecl}}$ ratio has no clear dependency on $M_{\mathrm{ecl}}$.
The B-DY channel contributes to the fraction with about $4\%$ and the S-DY channel has a contribution of about $1\%$.
This indicates that if the contribution to element enrichment from the mass loss of mergers is a constant fraction of $m_{\mathrm{m,tot}}$, then the total mass of polluting mass loss over the total cluster mass will also be a constant ratio.
The reason why $m_{\mathrm{m,tot}}/M_{\mathrm{ecl}}$ is constant will be explained in Section~\ref{sec:nm}.

\begin{figure}
  \includegraphics[width=1.0\columnwidth]{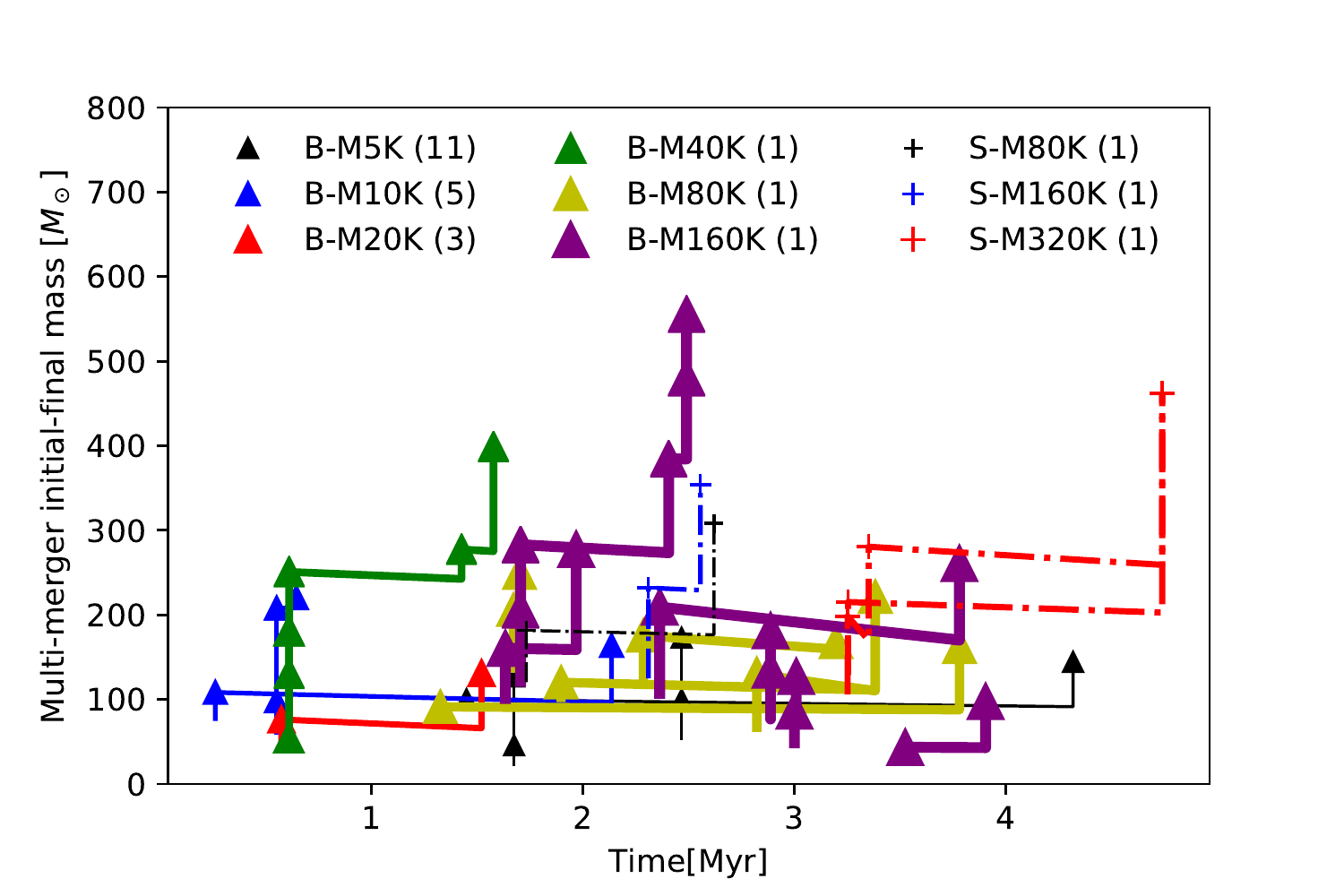}
  \caption{The merging history of multiple-merger stars.
    Different colors indicates different $M_{\mathrm{ecl}}$, and triangles (solid lines) are models with primordial binaries and crosses (dashed lines) are models with no primordial binaries.
    For each star, the initial mass and final mass ($m_{\mathrm{m}}$) at each merging time are shown with markers.
    The number in the bracket in the legend show the number of models with the same $M_{\mathrm{ecl}}$ (Table~\ref{tab:sets}).
  }
  \label{fig:mmulti}
\end{figure}

How the SMSs evolve is unclear due to the absence of observational constraints.
However, it is expected that a SMS has a short lifetime and a strong stellar wind \citep[e.g. a SMS may have significant mass loss around $1.5$~Myr, as shown in the model from ][]{Kohler2015}.
The \textsc{sse/bse} packages do not have a proper treatment of massive star evolution ($>150$~M$_\odot$), thus it is quite important to know the merging history of a SMS.
In Fig.~\ref{fig:mmulti}, we show all stars that have a multiple-merger history in all of our models.
In the B-M80K and B-M160K models, more than one massive star forms via multiple mergers.
In the B-M160K model, the most massive star with $555$~M$_\odot$ has suffered $5$ mergers within $1$~Myr.
The first $2$ mergers happen in a very short timescale ($<100$~yr of each other) as a chain of mergers.
The last $3$ occurred after $0.7$~Myr.
The multiple mergers of stars that produce a massive star ($>100$~M$_\odot$) from stars below $50$~M$_\odot$ within a short timescale is also found in the low-mass star cluster models done by \cite{Oh2018}.

\subsection{Mergers -- stellar mass and fast rotators}

\begin{figure}
  \includegraphics[width=1.0\columnwidth]{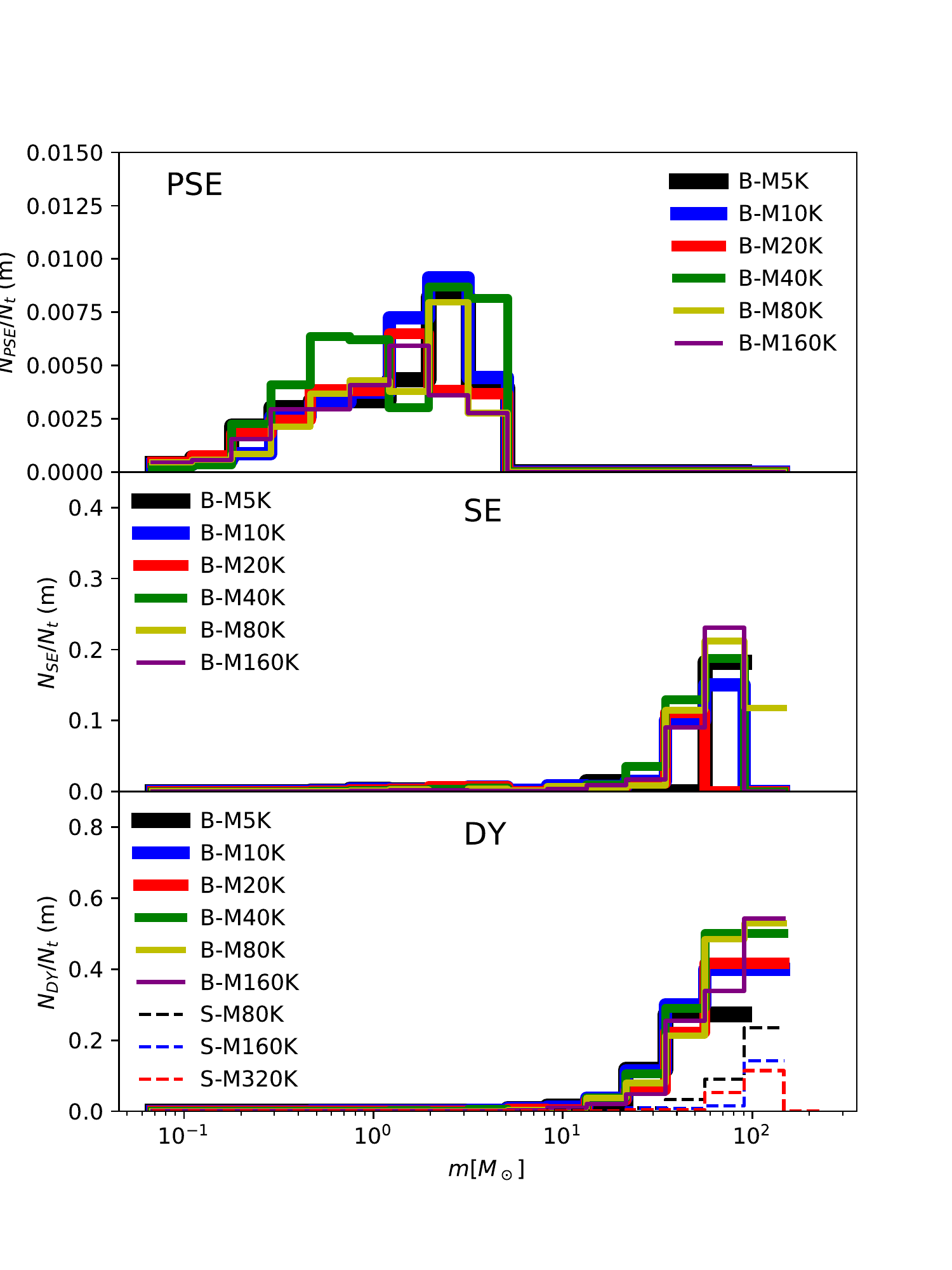}
  \caption{The number of stars involved in mergers v.s. the total number of stars in different mass regions with logarithmic bins.
    The mergers from different channels (PSE, SE and DY) are shown in separate panels.
    Different colors indicate different models.
  }
  \label{fig:nm}
\end{figure}

It is interesting to know how large the fraction of stars is in different mass regions that are involved in the mergers.
Mergers can also produce fast rotators if the binary orbital angular momentum can be transferred to the newly formed stars.
In Fig.~\ref{fig:nm}, we show the fraction of stars involved in mergers for different mass bins.
The different merger channels are also shown separately.
It is evident that the PSE mergers contribute mostly in the low-mass region ($m<5$~M$_\odot$), but fewer than $1\%$ of the low-mass stars suffer mergers.
According to \cite{Belloni2017}, although the period distribution of high-mass binaries is peaked around $1$~day, the eccentricity distribution is concentrated in the circular region.
On the other hand, for a canonical IMF, the number of high-mass binaries is much lower compared to the number of low-mass binaries.
These two aspects probably explain why the PSE mergers do not happen in the high-mass region.
A large fraction of high-mass stars experience mergers via both the SE and the DY channels.
Especially more than $50\%$ of the stars above $30$~M$_{\odot}$ have merged with others (many through the DY channels).
This is not very different for different B- models and it explains why the $m_{\mathrm{m,tot}}/M_{\mathrm{ecl}}$ ratio is constant.
Since almost all massive stars are a merger product and they dominate the total mass in mergers, $m_{\mathrm{m,tot}}$ is determined by the fraction of massive stars, in other words, determined by the IMF.
If the mergers are the major contributors for MSPs, the most massive OB mergers would be the dominant polluters.

\label{sec:nm}

\subsection{Evolution of the clusters}

\begin{figure}
  \includegraphics[width=1.0\columnwidth]{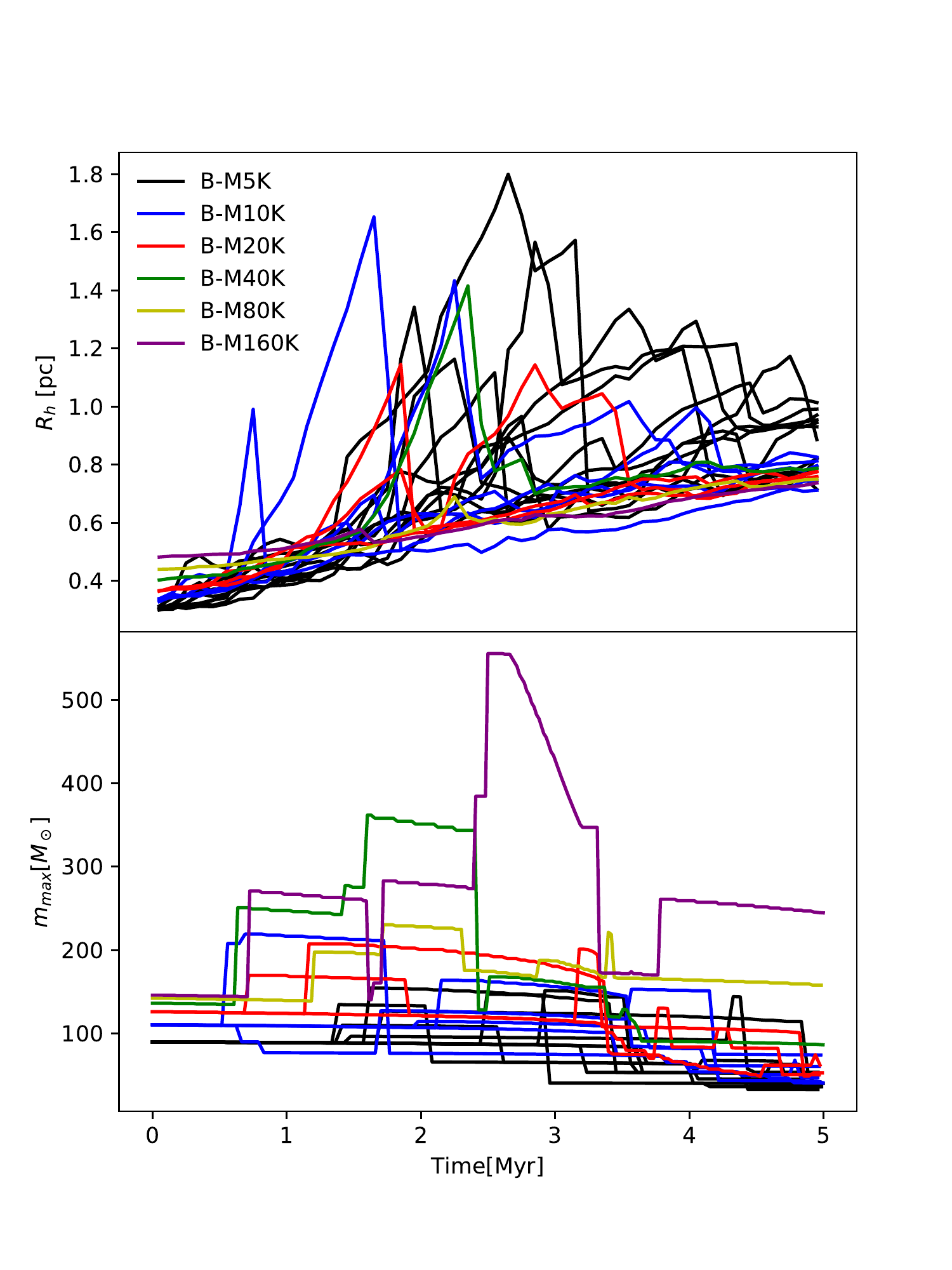}
  \caption{The evolution of half-mass radius (upper panel) and maximum mass of stars (lower panel) for models with primordial binaries.
  }
  \label{fig:rh}
\end{figure}

During the first $5$~Myr, the mass loss via stellar winds of massive stars drives the expansion of the star clusters in the present models which do not, by computational necessity, contain a star-forming gas content.
In the upper panel of Fig.~\ref{fig:rh}, we show the evolution up to $5$~Myr of the half-mass radius, $R_{\mathrm h}$, for all models with primordial binaries.
The increase of $R_{\mathrm h}$ for different models shows a consistent dependence on the time.
This suggests that stellar winds are the driving mechanism for the expansion, because the mass loss rate for stellar winds scales with the physical time but not the dynamical time.
In the realistic case where the cluster is still embedded, the expansion is likely to be smaller because the residual gas can slow down and capture the winds.
Thus even fast winds which exceed the escape velocity of clusters may well remain, mix with gas and contribute to the element enrichment.

There are also large and random jumps of $R_{\mathrm h}$.
This is caused by the ejection of massive stars via few-body interactions.
In our primordially mass-segregated models, massive stars dominate the total mass of the cluster center.
Once a massive star, especially the massive merger, moves outwards, $R_{\mathrm h}$ is perturbed.

In the lower panel of Fig.~\ref{fig:rh}, we show the evolution of the maximum mass of stars.
Due to the mergers, sudden jumps appear sometimes.
Then it smoothly decreases due to stellar winds or jumps down because of escape.
The maximum mass observed in our models is $555~$M$_\odot$ at about $2.4$ Myr (also shown in Fig.~\ref{fig:mmulti}), which decreases to about $350$~M$_\odot$ within about $1$~Myr due to a strong stellar wind.
Escapers and fast winds of massive stars prevent the growth of the stellar maximum mass via mergers.
Thus, the appearance and value of the maximum mass is highly stochastic (also shown in Fig.~\ref{fig:nmecl}).

\section{Multi-channel polluters}
\label{sec:polluters}

\subsection{The contribution from different polluters}

\begin{figure}
  \includegraphics[width=1.0\columnwidth]{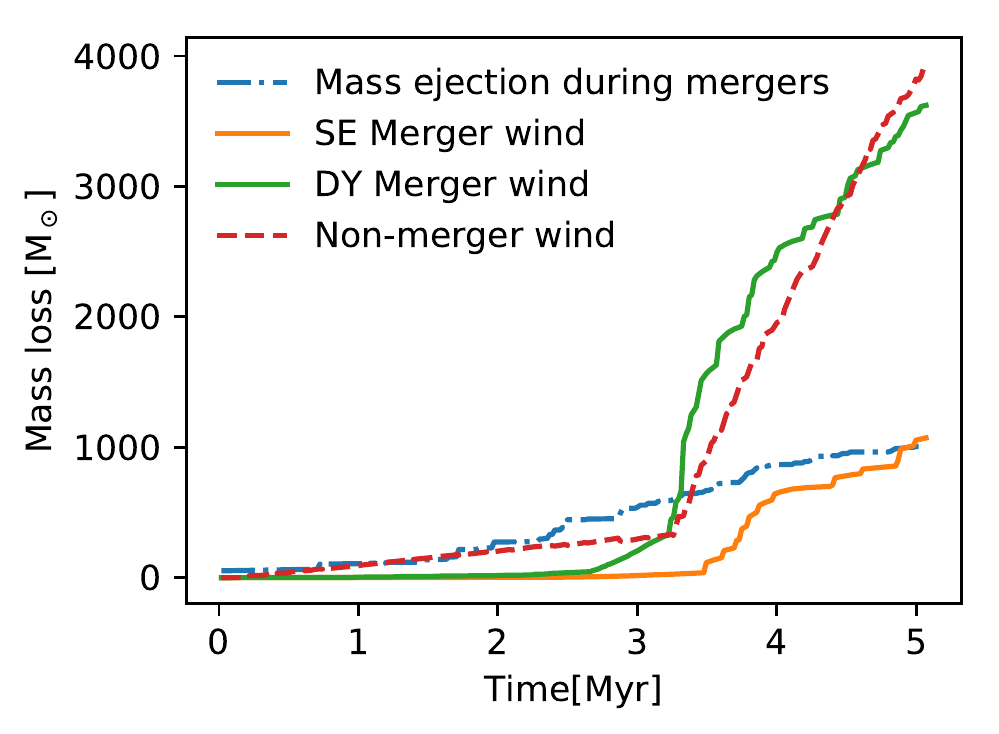}
  \caption{The cumulative stellar-wind mass loss from different channels in the model B-M160K.
    Mass ejection during mergers (BINARY channel) is estimated by assuming $30\%$ of the secondary mass in a binary is ejected during a merger.
    The SE/DY merger winds are the cumulative stellar wind mass loss from the new stars formed after mergers.
    Non-merger wind is the cumulative stellar wind mass loss from all stars that do not experience a merger.
  }
  \label{fig:wind}
\end{figure}

Above we have shown the properties of mergers in our $N$-body models.
In this section, we discuss how different scenarios might contribute to the formation of MSPs.
Fig.~\ref{fig:wind} shows the evolution of the cumulative stellar-wind mass loss from different channels in the model B-M160K.
We assume the mass ejection during a merger ($30\%$ of the secondary masses in binaries) represents the element-enrichment source from the BINARY channel.
The SE/DY Merger wind shows the cumulative stellar-wind mass loss from the new stars after mergers.
Since the new stars can be either FRMS or SMS, this type of mass loss represents the contribution from the sum of FRMS and SMS channels.
Our models do not produce the type of SMS discussed in \cite{Denissenkov2014,Prantzos2017,Gieles2018} due to our low-mass initial conditions compared to the massive GCs, thus the SMS contribution is not shown separately here.
Non-merger wind indicates the stellar-wind mass loss from stars that do not experience mergers, which is not considered as the source for the anti-correlation of Na-O and Al-Mg.

The result clearly indicates that the DY merger wind contributes about $40\%$ of the total stellar wind mass loss.
Thus, FRMSs and SMSs seem to be the dominant element-enrichment source.

\subsection{Binary channel}

In our models (Fig.~\ref{fig:nmecl}), $m_{\mathrm{m,tot}}/M_{\mathrm{ecl}}$, is below $5\%$ before $5$~Myr and is independent of $M_{\mathrm{ecl}}$.
Thus, we expect that the contribution from the BINARY channel is independent of $M_{\mathrm{ecl}}$ For the invariant IMF.
Fig.~\ref{fig:wind} indicates that this channel provides about $10^3$~M$_\odot$ of enriched winds, being $0.625\%$ of $M_{\mathrm{ecl}}$.

\cite{deMink2009} have studied the binaries with mass accretion from one star to another.
In our models, the major fraction of mergers are high-eccentric or hyperbolic face-on collisions, which is different from the mass accretion.
It is unknown how such collisions (driven by the DY channel) contribute to the element enrichment.
But it is expected that an energetic collision of massive stars may carry out the elements from the convective core to the surface, thus the ejection contains enriched elements.

\subsection{Fast rotator channel}

Our models indicate that more than $50\%$ of massive stars suffer mergers within $5$~Myr (Fig.~\ref{fig:nm}).
This is consistent with the mergers in the low-mass star cluster models of \cite{Oh2018}.
Thus, a large fraction of massive stars may become FRMSs or SMSs.
Ignoring SMSs, Fig.~\ref{fig:wind} suggests that the winds from FRMSs can provide a maximum mass of about $3.12\%$ of $M_{\mathrm{ecl}}$ to the gas for the canonical IMF.
Similarly as the BINARY channel, if the IMF is invariant, the total mass fraction of FRMSs is independent of $M_{\mathrm{ecl}}$.

\subsection{SMS channel}

Since a SMS tends to have a short lifetime \citep[$<1.5$~Myr; ][]{Kohler2015}.
To grow the mass of the SMS, multiple-mergers should occur in a sufficiently short time interval before a significant mass loss via stellar winds.
Indeed, our models show that the star more massive than $500$~M$_{\odot}$ suffered a chain of mergers with a maximum time interval of about $1$~Myr.
Thus the SMS formation is highly stochastic.
To estimate the contribution of SMS, we apply the fitting formula of Eq.~\ref{eq:mmax} to estimate the mean mass of a SMS in dependence of $M_{\mathrm{ecl}}$.
The mass loss of a SMS is significant when the mass is above $150$~M$_\odot$ as shown in \cite{Kohler2015}.
Thus we assume that the contribution of the stellar wind from a SMS is $M_{\mathrm{sms}}-150$~M$_{\odot}$, where $M_{\mathrm{sms}}$ is the mass of the SMS.

\begin{figure}
  \includegraphics[width=1.0\columnwidth]{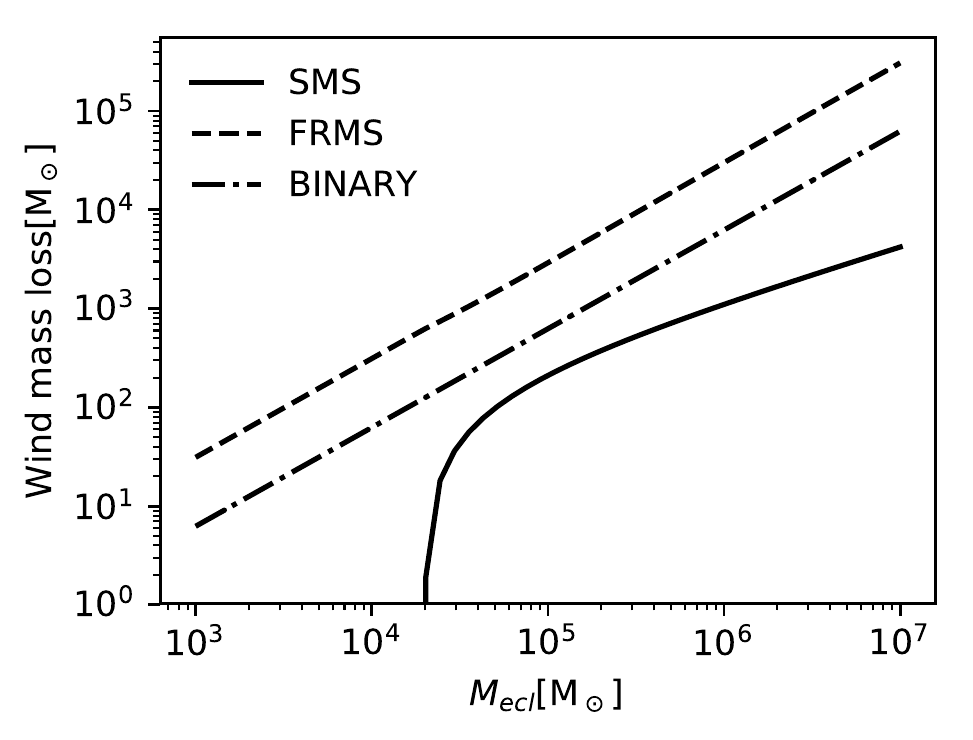}
  \caption{The cumulative stellar wind mass loss up to $5$~Myr from different MSP channels in dependence of $M_{\mathrm{ecl}}$.
    The wind mass via the SMS channel is $M_{\mathrm{sms}}-150$~M$_{\odot}$, where $M_{\mathrm{sms}}$ is calculated by using Eq.~\ref{eq:mmax}.
    The contribution from the FRMS channel is $3.12\%$ of $M_{\mathrm{ecl}}-(M_{\mathrm{sms}}-150~\mathrm{M}_{\odot})$ and from the BINARY channel it is $0.625\%$ of $M_{\mathrm{ecl}}$.
  }
  \label{fig:windest}
\end{figure}

In Fig.~\ref{fig:windest}, we show how the mean stellar-wind mass loss of a SMS depends on $M_{\mathrm{ecl}}$ and compare this with the expected wind contributions from the FRMS and the BINARY channels.
As shown by Eq.~\ref{eq:mmax}, the maximum mass of the SMS depends on $M_{\mathrm{ecl}}^{0.541}$, thus as $M_{\mathrm{ecl}}$ increases, the contribution to the total wind mass loss from the SMS channel becomes less than the FRMS and BINARY channels.
Since the scatter of $m_{\mathrm{m,max}}$ is large in our models, the extrapolation (Eq.~\ref{eq:mmax}) to large masses needs to be verified with massive $N$-body modelling, which is beyond current computational facility.
On the other hand, it is also possible that more than one SMS forms in a massive GC such that the contribution of the SMSs is underestimated in our models.

\cite{Gieles2018} predict much more massive SMSs compared to our scenario.
The mass of SMSs estimated by \cite{Gieles2018} is based on a semi-analytic model of runaway collisions without $N$-body simulations.
They also consider the mass growth of SMSs via gas accretion.
But primordial binaries are expected to heat the core of the cluster.
Thus the core density may not increase to the expected case predicted from the Gieles et al. scenario for a given size and a total mass of a cluster.
Thus it is unclear whether their prediction is real without detailed self-consistent $N$-body models.
On the other hand, our $N$-body models are relatively low-mass compared to their case and gas accretion is not considered.
Since runaway growth through collisions is an unstable process, the mass of the SMS may not have a simple scaling relation depending on the mass of the star cluster.
It is possible that our extrapolation underestimates the mass of SMSs in massive GCs.
This may be the reason for the different results of our models and those of Gieles et al.

\subsection{Variation of the IMF of the primordial population}
\label{sec:vimf}

In the scenarios that enriched populations form from the low-velocity stellar winds of polluters, IMFs should be very top heavy or even flat to solve the mass budget problem.
However, recent studies show that varying the IMF is limited by the constraints from the populations of BHs and the present-day morphology of GCs.
When the IMF is very top-heavy, the mass fraction of BHs increases significantly, and GCs will have a much shorter life time and the center surface brightness will be much smaller if a large number of BHs are still retained.
This has been confirmed by theoretical analysis, Monte-Carlo and $N$-body simulations \citep[e.g.][]{Breen2013,Chatterjee2017,Baumgardt2017,Giersz2019,Wang2019b}.

In our scenario, the total amount of the wind mass loss before $5$~Myr is about $10^4$~M$_\odot$, which is about $6.25\%$ of $M_{\mathrm{ecl}}$.
Mass loss from mergers contributes $60\%$ of the winds and is about $3.75\%$ of $M_{\mathrm{ecl}}$.
Although $3.75\%$ is a small fraction, when the winds are mixed with gas, a large number of young stars can still form from the enriched gas.
Thus the mass budget problem may not exist in our scenario, but whether the winds provide enough enriched elements to match the observed He, Na, O, Al, Mg abundance remains an open question (element budget issue).
Polluters in our scenario should produce more enriched elements per wind mass to match the same degree of element correlation as other scenarios predicted.
For example, if we assume the enriched populations to have half of their masses inheriting from the enriched winds and another half coming from the primordial gas, polluters should produce twice the enriched elements compared to the case where enriched populations form purely from the polluted winds.
Thus, to be consistent with observations, i.e, enriched populations contributing about $50\%$ of the total cluster mass, only $10-30\%$ of their mass can be from polluted winds.
This may require an abnormal stellar evolution model.
In our scenarios, we assume that all stars initially are not FRMS.
They form only via mergers.
It is possible that FRMS already exist in the primordial population.
Thus our estimated mass of the polluted winds becomes a lower boundary.
Assuming all stars are FRMS initially and all winds from the primordial population contributes the element enrichment, the mass of pollute winds increase to about $40\%$.
This may help to solve the element budget issue.

In our scenarios, varying the IMF also helps to solve the mass/element budget issue.
But since the primordial gas contributes to the mass of the enriched populations, we may not need extremely top-heavy IMFs.
In this work, we do not have a detailed element abundance analysis, but provide a rough estimation of how a top-heavy IMF may influence the fraction of the wind mass loss from polluters (up to 5~Myr) relative to the total mass at a given time (100~Myr here) of the enriched population ($f_{pwind}$), and the mass ratio between the young and the primordial population ($f_{msp}$) at this same time.
\cite{Marks2012b} suggested that the top-heavy of the IMF in a GC depends on its birth density and metallicity.
When the gas density is high and metal-poor, the IMF tend to be top-heavy.
Thus, it is possible that the primordial population in a GC had a top-heavy IMF, while the enriched populations may have a canonical IMF \citep{Kroupa2001}.
To be consistent with the survival of present-day GCs, a minimum value of the power index, $\alpha_3 = 1.5$, for the mass range of $1-150$~M$_\odot$ is applied (in this notation the Salpeter index is $\alpha_3 = 2.3$).
The escape of stars during the subsequent long-term evolution of GCs strongly depends on the internal dynamics and external galactic potential.
It is not easy to map our $5$~Myr models to the present-day GCs unless several uncertain assumptions are made.
To avoid this complexity, we ignore the escape of stars and only consider the wind mass loss up to $100$~Myr to estimate $f_{pwind}$ and $f_{msp}$.
After $100$~Myr, mass loss due to stellar winds becomes much weaker.
If the old and enriched populations are homogeneously mixed and the escape rate of all populations is the same, $f_{msp}$ remains the same during the long-term evolution of the GCs.
Observations show that enriched populations in many GCs tend to more concentrated radially, thus the escape rate of primordial population stars can be higher and our estimation becomes a lower boundary \citep[e.g.][]{Milone2009}.
Because all massive stars become remnants (BHs, NSs and WDs) after $100$~Myr, we only count luminous stars to calculate the total masses of old and enriched populations.
The wind mass loss is calculated based on the updated \textsc{sse} package \citep{Banerjee2019}.
We assume that the enriched populations have the same SFE ($30\%$) as the primordial population and the polluters contribute to the same fraction of wind mass loss ($60\%$) as in our models.
The latter assumption is valid when the merger rate of OB stars remains at about $50\%$, as shown in Fig.~\ref{fig:nm} and \cite{Oh2018}.

\begin{figure}
  \includegraphics[width=1.0\columnwidth]{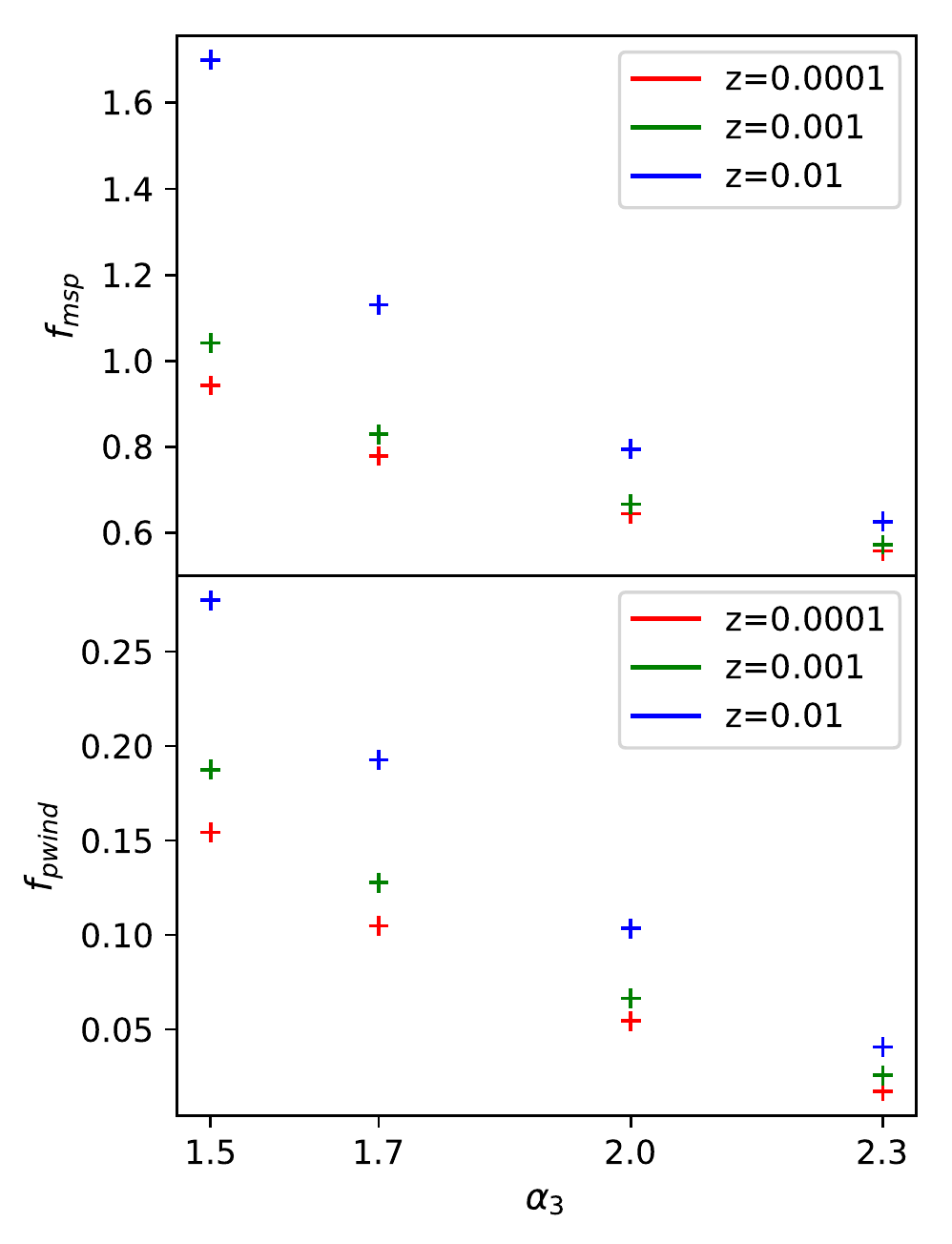}
  \caption{
    Upper panel: the mass ratio between enriched and primordial population (luminous stars) at $100$~Myr.
    Lower panel: the fraction of the wind mass loss from polluters (up to $5$~Myr) referring to the total mass of enriched populations at $100$~Myr.
    $\alpha_3$ of the IMF and metallicity for the primordial population are varied.
  }
  \label{fig:ratio}
\end{figure}

Fig.~\ref{fig:ratio} shows the results for $f_{\mathrm{pwind}}$ and $f_{\mathrm{msp}}$ depending on $\alpha_3$ and metallicity.
For the canonical Kroupa (2001) IMF, $f_{\mathrm{msp}}\approx 60\%$ and $f_{pwind} < 5\%$.
The values are not sensitive to metallicity.
When the metallicity increases and $\alpha_3$ decreases, both $f_{\mathrm{msp}}$ and $f_{\mathrm{pwind}}$ increase.
In the case of the maximally top-heavy IMF with $\alpha_3=1.5$,  $f_{\mathrm{msp}}$ and $f_{\mathrm{pwind}}$ show a large variation depending on the metallicity.
For $z=0.001$, $f_{\mathrm{msp}}\approx 1$, indicating that the two populations have similar masses.
The stellar wind from polluters contributes about $18\%$ for the mass of the enriched populations.

Here we assume the primordial and enriched populations have distinguishable IMFs.
When all populations form in a continuing star formation process, eventually no matter how many populations exist, they can be considered as one generation of stars forming with an age and metallicity spread.
The important point is that massive stars which become polluters should form relatively earlier than the major part of low mass stars in order to make the enriched populations fraction significant.
Thus, we can define the integrated IMF of all populations with an effective power index $\alpha_{\mathrm{3,eff}}$.
This $\alpha_{\mathrm{3,eff}}$ is larger than $\alpha_3$ for the IMF of the primordial population, like the analysis of the integrated galaxy initial mass function \citep[IGIMF theory, e.g.][]{Weidner2005,Jerabkova2018}.

\subsection{The lower mass limit of GCs with MSPs}
\label{sec:feedback}

Fig.~\ref{fig:ratio} shows that metal-rich GCs tend to have a larger fraction of the enriched populations, especially for the top-heavy IMFs.
The result is also independent of $M_{\mathrm{ecl}}$.
For low-mass clusters, $f_{\mathrm{msp}}$ can be large.
This is inconsistent with the observational data.
To explain why observed young and metal-rich star clusters do not have MSPs, we should also consider the energy feedback to the gas from the UV radiation, stellar winds and supernovae.
These feedback mechanisms are important to drive gas expulsion.

\cite{Krause2016} show that successful gas expulsion by winds and supernovae/hypernovae require a larger SFE when $M_{\mathrm{ecl}}/r_{\mathrm h}$ increases.
The higher SFE is needed as it implies a smaller gas mass needs to be removed from the cluster.
They also suggest that radiation feedback is not important.
Since SFEs and initial compactness are not well constrained in GCs yet, it is hard to draw a certain conclusion whether the feedback can successfully remove the gas.
\cite{Bailin2019} argue that the majority of GCs have non-zero but very small iron spread, and more luminous GCs (above $10^5$~$L_\odot$), which indicates larger $M_{\mathrm{ecl}}/r_{\mathrm h}$, have increasingly large iron spreads on average.
This suggests that the gas is probably removed successfully by a few supernovae/hypernovae with a time interval in most GCs.
It is possible that the star formation continued during this short interval thus a small iron spread exist.
GCs with larger $M_{\mathrm{ecl}}/r_{\mathrm h}$ require more time to have a complete gas expulsion, thus the iron spread is larger.

We carry out a similar analysis for the clusters with top-heavy IMFs.
First, we estimate the cumulative kinetic energy of stellar winds and supernova winds.
The stellar wind speed is estimated as the surface escape velocity of a star,
\begin{equation}
  v_{esc,wind} = \sqrt{\frac{2 G m}{r_{\mathrm s}}},
\end{equation}
where $m$ and $r_{\mathrm s}$ are the mass and radius of a star respectively.
The velocity of the supernova ejection (wind) is assumed to be a constant.
We use two velocities, $3000$ and $7000$~$km/s$, to represent the lower and upper boundaries.

Then we assume the gas has a Plummer profile with the same half-mass radius of the star clusters.
The Plummer potential energy of the gas is
\begin{equation}
  \Phi_{\mathrm{gas}} = -\frac{3\pi}{32 \sqrt{2^{2/3}-1}} \frac{G M_{\mathrm{gas}} (M_{\mathrm{gas}}+M_{\mathrm{ecl}})}{r_{\mathrm{h}}},
\end{equation}
where $M_{\mathrm{gas}} = M_{\mathrm{ecl}} \times (1-SFE)/SFE$ is the total mass of gas.

\begin{figure*}
  \includegraphics[width=1.0\textwidth]{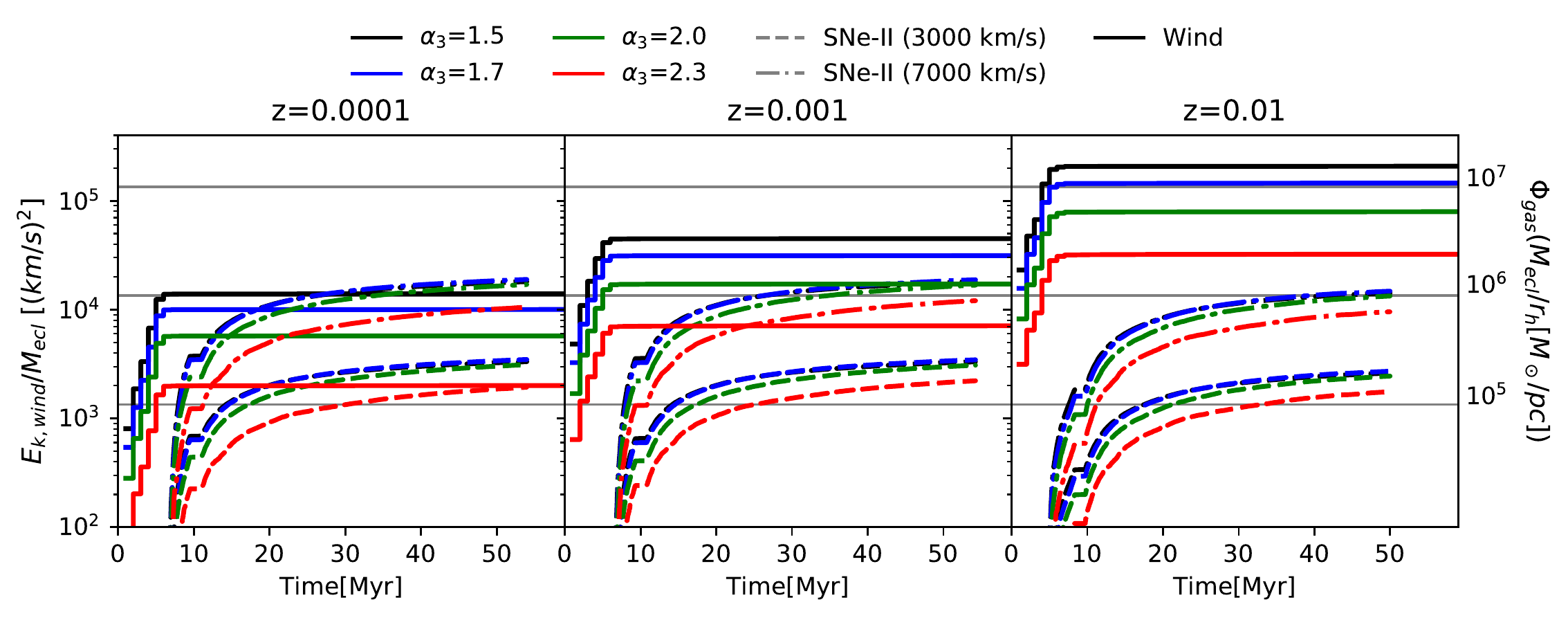}
  \caption{
    The evolution of the cumulative kinetic energy of stellar winds and supernova winds normalized by $M_{\mathrm{ecl}}$ for different IMFs ($\alpha_3$) and metallicity.
    Each panel shows the result of one value of metallicity.
    Different colors indicate $\alpha_3$ and line types represent wind types.
    The grey horizontal lines show three values of $M_{\mathrm{ecl}}/r_{\mathrm h}$, $10^7$, $10^6$ and $10^5$ ($\mathrm{M}_\odot/\mathrm{pc}$) from top to bottom.
  }
  \label{fig:ewind}
\end{figure*}

Fig.~\ref{fig:ewind} shows the evolution of cumulative kinetic energy for stellar winds and supernova winds, $E_{\mathrm{k,wind}}$, normalized by $M_{\mathrm{ecl}}$.
The grey lines show $\Phi_{\mathrm{gas}}$ for three values of $M_{\mathrm{ecl}}/r_{\mathrm h}$.
$E_{\mathrm{k,wind}}$ contributed from stellar winds increases rapidly in the first $5$~Myr and is almost constant later on, while the supernova wind contribution to $E_{\mathrm{k,wind}}$ happens later.
If we assume that all kinetic energy of winds contribute to push the gas away from the cluster and there is no transfer between the thermal energy and potential energy of the gas,
\begin{equation}
  \label{eq:gas}
  E_{\mathrm{k,wind}}(60 Myr) = \Phi_{\mathrm{gas}}
\end{equation}
provides the minimum boundary condition for complete gas expulsion.

For a given $M_{\mathrm{ecl}}/r_{\mathrm h}$ with a fixed SFE, top-heavy IMFs support the gas expulsion.
In the metal-rich star clusters ($z=0.01$), even for $M_{\mathrm{ecl}}/r_{\mathrm{h}}=10^6 \mathrm{M}_{\odot}/\mathrm{pc}$, stellar winds seem to be possible to completely remove the gas during the first $5$~Myr.
Thus, our scenario cannot work in metal-rich clusters since gas expulsion is too efficient such that new star formation is probably suppressed.
On the other hand, low-mass (density) star clusters also tend to be gas free due to the stellar winds.

\begin{figure*}
  \includegraphics[width=1.0\textwidth]{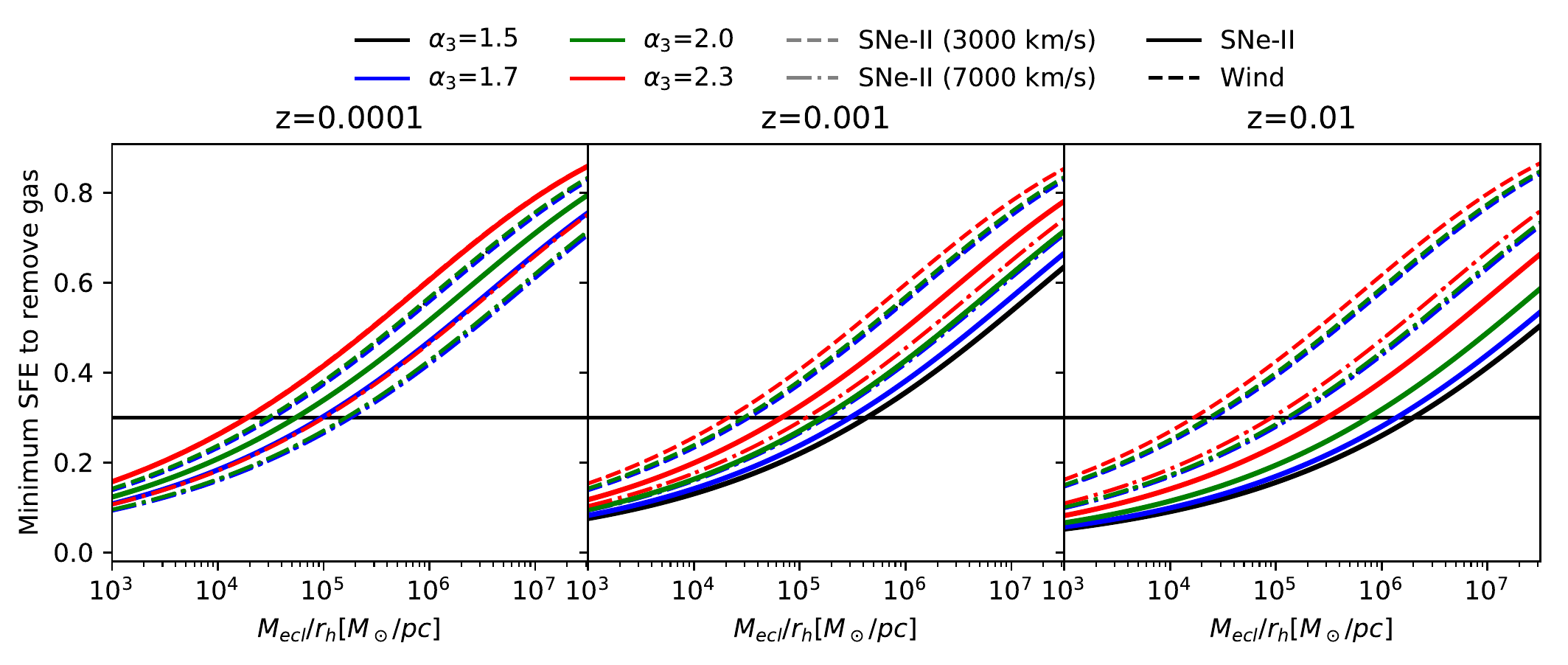}
  \caption{
    The relation between SFE and being $M_{\mathrm{ecl}}/r_{\mathrm h}$ when Eq.~\ref{eq:gas} is satisfied.
    The contribution from stellar winds and supernova winds to $E_{\mathrm{k,wind}}$ are calculated separately, represented by different line types.
    The plotting style is the same as Fig.~\ref{fig:ewind}.
    The black horizontal line shows SFE=$30\%$ for a typical star formation event.
  }
  \label{fig:sfe}
\end{figure*}

For a larger SFE, the $\Phi_{\mathrm{gas}}$ decreases due to a smaller gas mass and $E_{\mathrm{k,wind}}$ increases because of more massive stars.
Thus for any given $M_{\mathrm{ecl}}/r_{\mathrm h}$, we can find a SFE to satisfy the criterion of Eq.~\ref{eq:gas}.
We calculate this minimum SFE depending on $M_{\mathrm{ecl}}/r_{\mathrm h}$ in Fig.~\ref{fig:sfe}.
Above the curves, $E_{\mathrm{k,wind}}(60 Myr) > \Phi_{\mathrm{gas}}$.
For a typical SFE=$30\%$ and $z=0.0001$, star clusters with $M_{\mathrm{ecl}}/r_{\mathrm h}<2\times 10^4 \mathrm{M}_{\odot}/\mathrm{pc}$ are gas free due to stellar winds during the first $5$~Myr.
When $z=0.01$, the boundary shifts to $2\times 10^5 \mathrm{M}_{\odot}/\mathrm{pc}$.
If we assume the MSP cannot form due to gas expulsion based on this boundary,
the observational data shown in \cite{Bastian2018} may be well explained.

The observationally deduced Helium enrichment, $\Delta Y$, depends on the present-day GC mass \citep{Milone2017}, suggesting that the responsible polluters more easily form in massive GCs.
This is also consistent with our results, because gas in a more massive GC requires a longer time to be completely removed, thus young stars forming until gas evacuation cumulative a larger $\Delta Y$.

\section{Discussion}
\label{sec:discussion}

\subsection{Impacts of stellar evolution models}

In this work, we use the \textsc{sse/bse} stellar evolution package implemented in the \textsc{nbody6++} code.
Given that it is implemented into the existing publicly available $N$-body codes, we cannot easily (without a very significant programming effort) replace this package by others.
The merger criterion depends on the stellar radii of two stars.
When the orbital peri-center distance is below the sum of the radii of two stars, the merger is triggered.
For massive stars, especially for the SMS generated via multiple-mergers, the stellar evolution is highly uncertainty.
\cite{Spera2019} have compared the different stellar evolution codes and suggested that for massive stars up to $100$~M$_\odot$ before the end of the MS evolution phase, the \textsc{sse/bse} package provides a consistent result with the \textsc{mobse} and the \textsc{sevn} codes.
Above $100$~M$_\odot$, the mass-radial relation in the \textsc{sse/bse} package may be inaccurate but is not expected to be significantly different (within an order of magnitude). 

The peri-center distance, $R_{\mathrm p}$, of SE mergers, shown in the upper panel of Fig.~\ref{fig:types}, indicate this criterion.
It is shown that the dynamical mergers generally have $R_{\mathrm p}$ much smaller than the stellar radius.
Thus the uncertainty of the mass-radius relation of stars can influence the merger rate for SE mergers, but not significantly for the DYC mergers. But considering the probably larger uncertainty of primordial binary parameters, this effect is not likely to be a dominant one.

\subsection{Impact of initial conditions}

We have performed $N$-body modelling of star clusters up to an initial stellar mass of $1.6\times 10^5$~M$_\odot$ for the first $5$~Myr with the available initial conditions constrained by observational data.
The $R_{\mathrm{h,0}}$, primordial binary properties, IMF and initial mass-segregation are determined based on the assumption of a universal and self-regulated star formation process \citep{Kroupa2013,Belloni2017,Yan2017}.
A variety of independent observational evidence supports these assumptions \citep[e.g.][]{Kroupa2013,Weidner2013,Duchene2018,Plunkett2018}.
Besides, there is also observational evidence supporting the KSB primordial binary model \citep[e.g.][]{Leigh2015,Belloni2017,Duchene2018}.
\cite{Leigh2015} compared the observed binary fractions within and outside the half-mass radius of GCs and the results from their Monte-Carlo models.
They found that the KSB eigen-evolution model is more consistent with the observational data than the assumption that GCs had initially low binary fractions.
\cite{Belloni2017} validated this hypothesis by investigating the CMD of GCs.
They improved the KSB model based on the observed cumulative color distribution of GCs.
We apply this updated version in this work.

However, in the KSB formulation \citep[the version of][]{Belloni2017}, the low-mass binary properties are a derived empirical model from inverse dynamical population synthesis which corrects the observed properties for dynamical processing, while the high-mass binary properties are based on the observed already dynamically evolved OB star binary populations.
The detailed observational data of binaries in young star clusters show complicated properties based on the metallicity, masses of the stars, and current observations cannot provide the full information of binary properties for the whole parameter space \citep{Duchene2013}.
Moreover, high-order multiplicity (initial triples/quadruples) is not considered in the KSB model yet.

The binary MSP channel strongly depends on the choice of the primordial binary model.
Especially for the PSE and the SE, mergers are very uncertain.
The KSB model is currently the best we can use as it is most realistic by being derived from a wide range of observational data.
If we consider the existence of high-order multiplicities, the merging process is expected to be more frequent via the instability of these systems.

The comparison of the S- and B- models (Table~\ref{tab:sets}) shows that the formation of SMSs, assuming no significant mass loss within $1$~Myr, does not very sensitively depend on the primordial binary properties.
But we also assume a GC to be initially fully mass-segregated due to the requirement of self-regulated star formation.
Thus OB runaway mergers can happen immediately.
Otherwise the massive OB stars should first mass segregate to the cluster center on the timescale
\begin{equation}
  T_{\mathrm{ms}}(m) \propto \frac{\langle m\rangle}{m} T_{\mathrm{rh}} ,
\end{equation}
where $m$ is the stellar mass of the segregating star and $\langle m\rangle$ is the averaged stellar mass.
Considering OB stars with $20$~M$_\odot$, in the B-M160K model, $T_{\mathrm{ms}}(20~{\mathrm M}_\odot) \approx 1$~Myr, and in the S-M320K model, it is about $2$~Myr.
All stars above $20$~M$_\odot$ have a shorter $T_{\mathrm{ms}}$.
Therefore, it is expected that initial mass-segregation would not have a strong impact on the merging results for massive stars.
This is also the case for SMS formation, because the time intervals for multiple mergers is within $1$~Myr (Fig.~\ref{fig:mmulti}).

In this work we only investigate a metallicity of $Z=0.001$ and the mass loss of our models is limited by the stellar evolution models of the \textsc{sse/bse} package.
It is expected that a different metallicity can influence the stellar winds, especially for the SE mergers.
In Section~\ref{sec:vimf}, we have performed stand-alone calculations using a stellar evolution code to estimate the stellar wind contribution with different metallicity.
Ignoring the dynamical effects, the impact of metallicity on $f_{\mathrm{msp}}$ is not significant for the canonical IMF, but is significant for top-heavy IMFs.

\subsection{The discrete sub-populations}

\cite{Milone2015} show that the GC NGC~2808 has clearly identified discrete sub-populations, which means the star formation in GCs can happen several times or in spatially localized regions if the embedded cluster is not self-enriched homogeneously.
In our scenario, patchy (non-homogeneous) self-enrichment is expected due to the discrete nature of stellar mergers.
All the enriched populations form before gas expulsion to avoid pollution by Fe.
During this period, although the winds from mergers are the source for element enrichment, the radiation feedback from the ionization sources (massive MS stars) might prevent the star formation processes.
The notion of dynamical ejections of massive stars discussed in \cite{Kroupa2018,Wang2019}, might provide the idea to naturally explain the sub-populations.
Their studies suggest that the ejection of OB stars remove the ionization sources in the cluster center, thus new star formation can resume.
As discussed in Section~\ref{sec:feedback}, in massive and metal-poor GCs, stellar wind and radiation feedback is not sufficent to remove the gas thus enriched populations can form with enriched gas.
There should be a competition between the star formation process and the radiation/wind feedback.
Few-body interactions can strongly perturb the balance since massive stars can be ejected from the cluster center (Fig.~\ref{fig:rh}), which can reduce the feedback energy and also influence the element enrichment rate.
Thus, the formation of enriched populations can fluctuate due to the escapers.
In such way, sub-populations may be created.
While much more detailed research is needed to better understand the influence of metallicity, feedback, mergers and stellar dynamical ejection, the results shown here and this discussion suggest that these processes may be relevant for the MSP phenomenon.

\section{Conclusion}
\label{sec:conclusion}

The multiple stellar populations (MSP) discovered in GCs raise a major puzzle concerning their origin.
Many models have been developed.
Most of them discuss the origins solely via the stellar evolution modelling.
In this work, we investigate if and how mergers of massive stars in initially binary rich embedded very young proto-GCs might produce mass loss enriched by nuclear synthesis products which, when mixed with the embedding gas, might contribute to the MSP phenomenon if star formation continues within the forming GC for a few Myr.
We perform a series of direct $N$-body simulations of GCs with different initial total stellar masses and the initial conditions constrained from observational data of young star formation regions.
We found that a large fraction $>50\%$ of massive stars ($>30$~M$_\odot$) can merge within the first $5$~Myr of a GC's life before the first supernovae drive the residual gas out of the GC.
Mergers of binary star components are driven by both binary stellar evolution and dynamical perturbations.

Since mergers link BINARY, FRMS and SMS channels together, several polluters can work simultaneously to enrich the surrounding gas (before complete gas expulsion by supernovae) where enriched populations form.
Our results indicate that with the optimally sampled IMF of \cite{Kroupa2001}, about $5\%$ of the total cluster stellar mass, $M_{\mathrm{ecl}}$, is involved in mergers during the first $5$~Myr (this fraction being independent of $M_{\mathrm{ecl}}$).
By calculating the contribution of the winds from different channels, we find for the canonical IMF that winds from multiple polluters contribute $60\%$ of the total wind mass loss in the first $5$~Myr before complete gas expulsion, which corresponds to $3.75\%$ of $M_{\mathrm{ecl}}$.
The remaining $40\%$ comes from normal stellar winds.
This total wind mass is small but once the winds are mixed with the residual embedding gas, a significant enriched populations can form.
We find the FRMS, produced from merged binaries, to be the dominant contributor to the enriched wind.
With the extrapolation formula (Eq.~\ref{eq:fit} and Fig.~\ref{fig:nmecl}), we expect that a GC with $M_{\mathrm{ecl}}=10^6$~M$_\odot$ can produce a SMS weighting $10^3$~M$_\odot$ via multiple mergers within $1$~Myr.
Whether SMSs can form remains an open question and their masses are highly stochastic due to the chaotic few-body interactions.

If the star formation efficiency is $30\%$ for both the primordial (first-formed) and the enriched populations, the mass ratio between enriched and primordial population, $f_{\mathrm{msp}}$, after $100$~Myr depends on metallicity and the IMF of the primordial population, as shown in Fig.~\ref{fig:ratio}.
The mass budget problem does not seems to appear as a major problem.
When the primordial population has a top-heavy IMF and the enriched populations have a canonical IMF (which is a rough approximation for the metal-enriched IMF being more normal; \citealp{Marks2012b}\footnote{We follow the idea of \cite{Marks2012b} that the IMF tends to be top-heavy for a star formation region when the initial gas density is high and metal-poor, but do not apply their estimated $\alpha_3$ of IMF for GCs. Many GCs in their samples are estimated to have extremely top-heavy IMF with $\alpha_3 \approx 0.8-1.5$, due to the absence of the dynamical effect of BHs in their analysis.}), $f_{\mathrm{msp}}$ increases and the fraction of enriched elements in the enriched populations also increases.
Thus varying the IMF can help to match the element requirement constrained by the observational data.
However, our scenario does not need an extremely top-heavy (flat) IMF required by the scenarios where enriched populations form from the low-velocity winds (e.g. the AGB scenario).
Present-day GCs are unlikely to have had extremely top-heavy IMF due to the constraint from long-term dynamical evolution (especially the impact on cluster expansion from BH subsystems; \citealp{Breen2013,Chatterjee2017,Baumgardt2017,Giersz2019,Wang2019b}).
Fig.~\ref{fig:ratio} suggests that to reach $f_{\mathrm{msp}} \approx 0.5-1.0$, $\alpha_3 \approx 1.5 - 2.3$, assuming both populations have the same fraction of escapers during the long-term evolution.
But if the enriched populations form more centrally concentrated, the primordial population are preferentially removed from the cluster via tidal evaporation.
Therefore, $f_{\mathrm{msp}}$ increases during the long-term evolution of GCs.
Notice here $\alpha_3$ is for the IMF of the primordial population, the effective (average) $\alpha_{\mathrm{3,eff}}$ for the integrated IMF of all populations have a relative larger value \citep[like the case of IGIMF theory, e.g.][]{Weidner2005,Jerabkova2018}.

The rough estimation of the stellar wind feedback energy that can push the gas away from the clusters, when compared to the gas potential energy, are shown in Fig.~\ref{fig:windest} and \ref{fig:sfe}.
The results suggest that with a typical SFE$=30\%$, low-mass and metal-rich star clusters are unlikely to have MSPs, which is consistent with the observational data.
On the other hand, massive stellar escapers driven by few-body interactions also cause the fluctuation of feedback, which can result in discrete sub-populations.
Discrete sub-populations may also appear due to inhomogeneous self-enrichment of the embedding star-forming gas due to the discrete merger events.

Due to software and hardware limitations, this work has not included detailed stellar evolution calculations to clarify the enrichment of each different element in comparison with observational data.
This needs to be addressed in the future with the improved $N$-body method and stellar evolution packages, which are able to take into account million stars with a large number of primordial binaries and gas self-consistently.
Given the severe computational costs, we chose to investigate sets of parameters (radius, mass, binary population, IMF) which are well discussed in the literature.
It is beyond the scope of this work to attempt different (and even random) combinations.
This work seeks to suggest that stellar mergers in initially binary-rich clusters may play a role in the elemental abundance peculiarities observed in GCs (by linking multiple MSP scenarios together).
Thus we use previously published distribution functions (for stellar masses, binary star properties, cluster radii), that is, we do not adapt these to achieve a particular result.

We acknowledge that there are major uncertainties on, e.g., the amount of mass lost from the stars when they merge and which layers of the star are ejected.
The liberation of elements from the merger depends on the exact merger mechanics and on the evolved state of the merging stars.
Thus, mergers in the first few $10^5$ years are likely to introduce different combinations of elements than mergers during the later times before $5$~Myr when the merging stars are already evolved.
Mergers of stars formed from previously enriched material (through mergers and winds) are likely to contribute yet different combinations of elements.

With this work we are merely suggesting that mergers in initially binary rich embedded massive clusters may be playing a potentially very important role for the elemental abundances of stars which continue forming throughout the cluster while the dynamical activity is on going and before the gas is fully blown out by (most probably) supernovae.

\section*{Acknowledgments}
L.W. thanks the Alexander von Humboldt Foundation for funding this research.
We thank Sambaran Banerjee for useful discussions.

\label{lastpage}

\end{document}